\documentclass[twocolumn,floats,pra,showpacs]{revtex4-1}

\usepackage{amsmath}
\usepackage{amsfonts}

\usepackage{tikz}

\newcommand {\be}{\begin{equation}}
\newcommand {\ee} {\end{equation}}
\newcommand {\bea}{\begin{eqnarray}}
\newcommand {\eea} {\end{eqnarray}}

\newcommand{\bk}{{\bf k}}

\newcommand{\bx}{{\bf x}}
\newcommand{\hati}{{\hat{\bf i}}}
\newcommand{\hatj}{{\hat{\bf j}}}
\newcommand{\hatk}{{\hat{\bf k}}}

\newcommand{\kket}[1]{\left| #1  \right>}

\begin{document}


\title{Compactly-supported Wannier functions and algebraic $K$-theory}
\author{N. Read}
\affiliation{Department of Physics, Yale University, P.O. Box 208120, New Haven, CT 06520-8120, USA}
\date{February 24, 2017}

\begin{abstract}
In a tight-binding lattice model with $n$ orbitals (single-particle states) per site, Wannier functions
are $n$-component vector functions of position that fall off rapidly away from some location, and such that
a set of them in some sense span all states in a given energy band or set of bands; compactly-supported
Wannier functions are such functions that vanish outside a bounded region. They arise not only in band
theory, but also
in connection with tensor-network states for non-interacting fermion systems, and for flat-band
Hamiltonians with strictly short-range hopping matrix elements. In earlier work, it was proved that for
general complex band structures (vector bundles) or general complex Hamiltonians---that is, class A
in the ten-fold classification of Hamiltonians and band structures---a set of compactly-supported Wannier
functions
can span the vector bundle only if the bundle is topologically trivial, in any dimension $d$ of space, even
when use of an overcomplete set of such functions is permitted. This implied that, for a free-fermion
tensor network state with a non-trivial bundle in class A, any strictly short-range parent Hamiltonian
must be gapless. Here, this result is extended to all ten symmetry classes of band structures without
additional crystallographic symmetries, with the result that in general the non-trivial bundles that
can arise from compactly-supported Wannier-type functions are those that may possess, in each of $d$
directions, the non-trivial winding that can occur in the same symmetry class in one dimension, but nothing
else. The results are obtained from a very natural usage of algebraic $K$-theory, based on a ring of
polynomials in $e^{\pm ik_x}$, $e^{\pm ik_y}$, \ldots, which occur as entries in the Fourier-transformed
Wannier functions.
\end{abstract}


\maketitle

\section{Introduction}
\label{intro}

The subject of topological phases in quantum non-interacting particle systems, or in linear wave-equation
systems, has grown into a major area of research in condensed matter physics, which includes the free
(i.e.\ non-interacting) fermion approximation to topological insulators and superconductors. Various
approaches to the latter problems have lately been converging around questions of which of the phases can
be represented by examples that possess sets of compactly-supported wave packets for a single particle,
constructed from states in a single band or from a subset of the bands in $\bf k$ space, that are in some
sense complete sets (like Wannier functions), and with associated single-particle Hamiltonians in which
the hopping matrix elements
are strictly short range (i.e.\ their range is bounded). (We consider only systems that are invariant under
a discrete group of translations on a lattice in real [position] space, and have in mind tight-binding
models that have only a finite-dimensional space of orbitals [single-particle states] available at each
lattice site.) The various approaches just mentioned are (i) tensor network states \cite{dubail,cirac}---the
extension of matrix product states \cite{Cirac_MPS} to more than one dimension---which can be applied
to interacting \cite{PEPS} as well as to
non-interacting particles; (ii) compactly-supported Wannier-type functions \cite{ozolins,budich},
an extreme example of Wannier functions \cite{wannier}, of interest in electronic structure calculations;
and (iii) flat-band Hamiltonians with strictly short-range hopping \cite{seidel}, an extreme form of the
flat-band approach popular in constructions of topological insulator states, including ones with
interactions.

In each of these areas, which have been progressing largely independently, there are by now
``no-go'' theorems \cite{seidel,dubail} and numerical results \cite{budich} that in each case say
(expressing it loosely) that some topologically non-trivial
phases cannot be constructed with the techniques mentioned (and with an appropriate gap in the spectrum
of the Hamiltonian in the model) due to some sort of obstruction. The cases ruled out are some of those
occurring in a space of dimension $d$ larger than $1$. On the other hand, constructions as matrix-product
states have long been known for some non-trivial topological phases in dimension $d=1$ (even if not
always under those names), including some for non-interacting fermions \cite{KitaevChain}. But in each
approach, many cases among the distinct non-trivial topological phases remained unresolved.

In this paper we provide a unified view of these techniques, and a full solution of the problem for
non-interacting topological phases in each of the ten symmetry classes \cite{az,srfl,kitaev} that arise
on a lattice
(in a tight-binding model) that possesses translation symmetry, but no other crystallographic symmetry,
and in all dimensions of space. The essential nature of the problem lies in the form of functions in
$\bk$ space, which are vector functions of $\bk$ with entries that are polynomials in $e^{\pm ik_x}$,
$e^{\pm ik_y}$ and so on, that lie in a band (or in the span of the states in a set of bands) for all
$\bk$. In position space, these become packets that have compact support, that is, they vanish outside
some bounded region of the lattice. (Single-particle tight-binding Hamiltonians that
are strictly short-range have matrix elements that are the same type of polynomials, when written in
$\bk$ space.) These polynomial functions in $\bk$ space form certain algebraic rings, and lead to the use of
algebraic methods. The
solution to the question of which phases can be constructed (subject to some conditions, and under a
certain notion of equivalence of topological phases) is given by a classification that uses
{\em algebraic} $K$-theory of the given rings, in contrast to the topological $K$-theory \cite{atiyahbook}
now familiar
to physicists in the classification of non-interacting topological phases in general \cite{kitaev}.
The solution reveals that the only non-trivial topological
phases that can be constructed in these ways in dimensions higher than $1$ are those that are non-trivial
only because they utilize topology that comes from the one-dimensional case in the same symmetry class,
applied to each of the $d$ directions in space, together with topology that comes from the 
zero-dimensional case in the same symmetry class; thus for $d>1$ these are particular examples of ``weak''
topological insulators and superconductors. We also comment on the form of this general result, and
speculate that something similar should hold for interacting tensor network states.
This work was motivated by the necessity of extending the previous results on tensor network states by
Dubail and the author \cite{dubail} (to be referred to as DR) to other symmetry classes.

In the remainder of this introduction, we review the general problems,
the older results, and work by others within the various approaches, then describe our
results. Readers should note that some basic terms used later are defined in this section only.

\subsection{Compactly-supported Wannier-type functions}

The basic natural place at which to begin is the compactly-supported Wannier-type functions, which are
the central
topic of the discussion, and which will be defined in this Section. First, however, we recall the nature of
energy-band structure. For a tight-binding model on a lattice (with the translation-symmetry assumption
mentioned above), the single-particle Hamiltonian is a matrix with rows and columns labeled by pairs
$(i,\alpha)$, where $i$ denotes a site in a Bravais lattice and $\alpha=1$, \ldots, $n$ labels the
single particle states, or orbitals, on each lattice site; $\alpha$ could subsume any spin indices.
(Models for other, non-Bravais, lattices can be brought to this form by grouping lattice sites in 
the same unit cell together onto sites of a Bravais lattice, and treating
each group as a single site.) For the present, the Hamiltonian is generic; it is not required to be real,
nor to have any symmetries other than under translations. In Fourier (or Bloch) space, single particle
states are labeled instead by $\bk$ and $\alpha$ ($\bk$ has components $k_\mu$, $\mu=1$, \ldots, $d$),
with $\bk$ in the Brillouin zone which, because of equivalence of $\bk$s that differ by addition of
reciprocal lattice vectors, topologically is a $d$-dimensional torus, which we call
$T^d$. The Hamiltonian in $\bk$-space is diagonal in $\bk$, and so is an $n\times n$ matrix function 
of $\bk$. Its energy
eigenvalues form a set of $n$ continuous real functions of $\bk$, called energy
bands. (The bands may cross, making their identification as $n$ distinct functions non-unique.) We are more
interested in the corresponding eigenvectors, each of which is an $n$-component vector that varies
with $\bk$. Then the eigenvectors corresponding to any subset of $m$ of the bands at any $\bk$ span
a subspace of dimension $m$ in the $n$-dimensional vector space.

Wannier functions have a long history; see Refs.\ \cite{wannier,kohn}. They may be associated with 
a single band, or with a set of bands.
A Wannier function is a wavepacket in position space, taking values in the space spanned by the orbitals
for each site; thus it is a vector with components labeled by the indices $i$, $\alpha$. In the original
definition, it and its translates on the lattice are further supposed to be constructed from the states
in a single band, and to be orthogonal to one another. This can be accomplished if each of them is
the inverse Fourier transform of a single function of $\bk$ that is a normalized eigenvector in the
desired band at each $\bk$. The lattice translations of such a Wannier function correspond to inverse
transforms of the same function of $\bk$ multiplied by integer powers of $e^{ik_\mu}$ for each
$\mu$; the power in each  $e^{ik_\mu}$ determines the translation
of the function. (Here and below, for simplicity we treat the lattice as square, cubic,
or hypercubic; other Bravais lattices behave similarly, and are included by using non-orthogonal 
coordinates that correspond to the primitive translations of the lattice, while $\bk$ vectors are viewed 
as being in the dual space to these coordinates, so that no metric on space or reciprocal space is ever 
used.) More generally, one could consider a set of $m$ functions in $\bk$ space that are in
the span of the eigenvectors for a set of $m$ bands, vary continuously with $\bk$, and are orthonormal
at each $\bk$ in $T^d$. In order to make the Wannier functions well localized in position space, the
functions in $\bk$ space must be smooth, not just continuous. In recent years, there has been interest
in making the Wannier functions as localized as possible, in some definite sense; these are called
maximally-localized Wannier functions. See Ref.\ \cite{max_wannier} for a recent review.

We will define a broader notion. For theoretical purposes, orthonormality of the vector-valued functions
in $\bk$ space is not really required; one may consider only linear independence and completeness at
each $\bk$. In fact, in some situations, linear independence is not essential either, and we can drop it,
in particular we can allow more than $m$ functions of $\bk$ (still in the span of the same $m$ bands).
But completeness does seem important. Hence (following DR) we define a collection of
{\em Wannier-type functions} to be a set of continuous vector-valued functions of $\bk$ which,
at each $\bk$, lie in and span the $m$-dimensional subspace spanned by the eigenvectors in the $m$ bands in
question (the inverse Fourier transforms of these, and translations thereof, give the actual Wannier-type
functions in position space). We will see that the set can always be taken to be finite.

The language of vector bundles can be useful in these problems, even for band theorists. A
vector bundle
\cite{atiyahbook,milstash}
consists of a ``base'' space $B$, which for band theory is just the Brillouin torus $T^d$ (points in which
can be labeled by $\bk$), and for each $\bk$ a vector space of the same dimension, $m$ say, for all
$\bk$. (In our discussion, the vector spaces are complex.) The totality of vectors in these spaces forms 
the ``total space'' of the vector bundle, while
the vector space at each $\bk$ is called the ``fibre'' at $\bk$. The dimension of the fibre at each point
is called the ``rank'' of the vector bundle; we emphasize that (because the Brillouin torus is a connected
space)
it is the same at all $\bk$. In band theory for a tight-binding model, there is a rank $n$
vector bundle which includes all possible states in $\bk$ space. For a set of $m$ bands, the states in those
bands (may) form a rank $m$ vector bundle that is a sub-bundle of that; the fibre at each $\bk$ is a
subspace of the $n$-dimensional space. (The notation $n$ and $m$ for these numbers will be used fairly
consistently throughout the paper.) In order to be a vector bundle, it is crucial that the fibres vary
continuously with $\bk$;
here that means that the rank-$m$ subspace (or one could say, the projection operator onto the
subspace) varies continuously with $\bk$. (When bands cross, this may not be satisfied, depending on
the choice of which $m$ bands to consider, and then one does not obtain a vector bundle. But for a set of
bands
that occupy a range of energies and are separated at every $\bk$ by a gap or gaps from other bands higher
or lower in energy, this will hold.) Next, a ``section'' of a vector bundle is a function of $\bk$ that
takes values in the fibre of the vector bundle at each $\bk$, that is defined for all $\bk$, and continuous
in $\bk$. We can see that a Wannier function corresponds to a section of a vector bundle, however in general
sections are allowed to vanish at some $\bk$. One can consider sections, or sets of sections, that have
additional properties, such as being smooth, or
normalized, and so on. The virtue of the language of vector bundles in general is that there are situations
in which it is more convenient to speak of the fibre at $\bk$ as a vector subspace, rather than of
particular vectors, or of the totality of vectors at all $\bk$, rather than of particular sections.
Then in the language of vector bundles, Wannier-type functions correspond to a set of sections that
span the fibre of the vector bundle at every $\bk$. Even if the vector bundle is not given in advance, we
can define a
set of Wannier-type functions to be a finite set of continuous $n$-component vector functions of $\bk$
that at every $\bk$ span a subspace of dimension precisely $m$ (the same $m$ at all $\bk$); this then
defines the vector bundle.

Wannier functions are usually supposed to be well localized in position space. The inverse Fourier transform
of a section of an arbitrary vector bundle over the Brillouin torus may not be well localized, because a
section is only required to be continuous (as an $n$-component vector function of $\bk$). In order to obtain
Wannier functions that fall exponentially, asymptotically at large distance from the Wannier ``center'',
one requires that the section of the vector bundle be (real) analytic as a function of $\bk$ (again, this
means each component of the vector is an analytic function of $\bk$). We will sometimes use the term {\em 
analytic} for a rank-$m$ vector bundle that is a sub-bundle in the tight-binding model, if it has the 
property that, for every $\bk$ in the Brillouin torus, there is a set of $m$ sections of the vector 
bundle that are both (real) analytic and span the fibre at $\bk$, or equivalently if the projection 
operator into the fibre varies real-analytically with $\bk$ at all (real) $\bk$. Again, not all vector 
bundles satisfy this property, however, if there is an energy gap (above or below) separating the bands 
making up the vector bundle from the remaining bands at every $\bk$, then the vector bundle will be
analytic.

It has long been realized that in a topologically non-trivial band, Wannier functions in the
traditional sense do not decay rapidly with distance \cite{no_wannier}. This is typically described by
saying something like ``there is no smooth gauge'' for the function in $\bk$ space (the Fourier transform
of the Wannier function). When expressed in the language of vector bundles, this result becomes immediate.
First, we introduce the standard definition of a {\em trivial} vector bundle: a rank $m$ vector bundle is
(topologically) trivial if it has a set of $m$ sections that are linearly independent at all $\bk$ (in 
particular, none of them vanishes anywhere). Otherwise, of course, it is termed non-trivial. Note that 
this definition of trivial and non-trivial does not require examination of any Chern numbers and so on, 
as used for example in Refs.\ \cite{no_wannier}; Chern numbers, though possibly convenient computationally, 
in general give only partial characterizations of vector bundles anyway (i.e.\ a trivial vector bundle 
has all Chern classes zero, but the converse does not have to hold). (However, in low dimensions, 
i.e.\ $d\leq3$, the familiar first Chern numbers do characterize complex vector bundles up to 
isomorphism; see Sec.\ \ref{syzygy} below.)
Then for a non-trivial vector bundle, any attempt to find a linearly-independent set of $m$ sections 
(corresponding to a ``choice of gauge'')
will find that they cannot be made continuous (let alone analytic) at all $\bk$, while use of discontinuous 
``pseudo-sections'' will produce slowly-decaying tails in position space. (A similar result has been 
discovered for topological insulators \cite{sv}.)

Our definition of Wannier-type functions reflects an attempt to circumvent this result. Even a non-trivial
vector bundle can have a larger set of more than $m$ sections that span the fibre at every $\bk$,
but necessarily any subset of size $m$ becomes linearly dependent at some $\bk$. If the sections in the
set are analytic, then the corresponding Wannier type functions will decay rapidly, and may be useful, at
the cost of having to work with an ``overcomplete'' set of more than $m$ functions. Indeed, the
construction of a pair of time-reversal non-invariant sections of the occupied band bundle in a topological
insulator in Ref.\ \cite{sv} can be viewed as an example of this, if one includes the time-reversed
partners of the sections in the set (compare Section \ref{aiaii} below).

One type of highly-localized behavior for functions is that they could be {\em compactly supported}, 
that is, vanish outside some bounded region on the lattice in position space (this is not necessarily 
equivalent to other definitions of maximally localized). In Fourier space, such functions
become polynomials in $e^{\pm ik_\mu}$, that is the degree in each $e^{ik_\mu}$ is bounded both above
and below. We define
\be
X_\mu=e^{ik_\mu}
\ee
(for the hypercubic lattices); polynomials with both positive and negative powers of the variables
$X_\mu$ are called {\em Laurent polynomials}, while the usual kind with only non-negative
powers in $X_\mu$ will be called ``ordinary'' polynomials. (Frequently, the distinction is not
significant.) The use of such functions has recently been advocated and connected
with compressed sensing \cite{ozolins}. For band structure, the corresponding sections of a vector bundle
can be called (following DR) {\em polynomial sections}. It may be unlikely that a generic band structure 
has a vector bundle that admits polynomial sections. But for studies of model systems, one can consider 
band structures that have (over-)complete sets of polynomial sections for the vector bundle for, say, 
the filled bands---in other words, compactly-supported Wannier-type functions. (Such a vector bundle, 
which we will term {\em polynomially generated} in Sec.\ \ref{bunmod} below,
is necessarily analytic \cite{dubail}.) Then the question has been raised of whether such models exist 
for topologically non-trivial vector bundles (phases of matter) \cite{budich}. This is the problem that 
is solved in the present paper.

\subsection{Free-fermion tensor networks, parent Hamiltonians, and no-go theorem}
\label{sec:nogo}

A little earlier, similar issues were discussed in an apparently different setting, that of tensor network
states (TNSs). TNSs are a broad subject, but here we will describe only the free-fermion versions.

A ground state that corresponds to band structure of the sort we have been discussing, and with $m$
bands filled, has the general form in terms of second quantization
\begin{equation}
	 \exp \left( \int \frac{d^d {\bf k}}{(2\pi)^d} \, \sum_{\alpha,\overline{\alpha}}g_{{\bf
k},\overline{\alpha}\alpha}
\, c_{{\bf k},\overline{\alpha}}^\dagger c_{{\bf k},\alpha}  \right) \kket{11\cdots,00\cdots0}.
\label{genTNS}
\end{equation}
Here $\alpha=1$, \ldots, $m$, $\overline{\alpha}=m+1$, \ldots, $n$, and the reference state
$\kket{11\cdots,00\cdots0}$ is annihilated by $c_{\bk,\alpha}^\dagger$ ($c_{\bk,\overline{\alpha}}$)
for all $\bk$, or equivalently by $c_{\bx,\alpha}^\dagger$ ($c_{\bx,\overline{\alpha}}$) for all $\bx$,
and for all $\alpha$ ($\overline{\alpha}$).
The creation and annihilation operators obey
$\{c_{\bk\alpha},c_{\bk'\alpha'}^\dagger\}=(2\pi)^d\delta(\bk-\bk')\delta_{\alpha\alpha'}$
for $\alpha$, $\alpha'=1$, \ldots, $n$.
We write $g_{\bk,\overline{\alpha}\alpha}$ as $g_{\bf k}$, which is an $(n-m)\times m$ matrix of functions
of $\bk$ in the Brillouin zone, say $[-\pi,\pi]^d$ for the hypercubic lattice.

The ground state is annihilated by single-particle operators of the form
\be
d_{\bk}^\dagger=\sum_{\alpha} u_{\bk,\alpha}
c^\dagger_{\bk,\alpha}+\sum_{\overline{\alpha}}v_{\bk,\overline{\alpha}}
c^\dagger_{\bk,\overline{\alpha}},\label{eq:d-k}
\ee
where the coefficients obey
\be
v_\bk =g_\bk u_\bk,
\label{v=ug}
\ee
where $u_\bk$ ($v_\bk$) is an $m$-component ($n-m$-component) column vector.
These are creation operators for particles in states in the filled band, and so annihilate the
ground state. There are other operators of a similar form for the empty band.

For a TNS, the coefficients $g_{\bk \overline{\alpha}\alpha}$ must be ratios of polynomials
\cite{dubail,cirac} (if one $g_{\bk \overline{\alpha}\alpha}$ is initially a ratio of Laurent
polynomials, then
it can be turned into a ratio of ordinary polynomials by multiplying numerator and denominator by positive
powers of some $X_\mu$s). Then we can find solutions for $u$, $v$ as polynomials. The equations for them can
be rewritten with polynomial coefficients by multiplying each component by the lowest common denominator
in that row of $g$. Then they have the form
\be
Z_\bk w_\bk=0,
\label{poly_eq}
\ee
where $Z_\bk$ is fixed $(n-m)\times n$ matrix with polynomial entries and $w_\bk$ is an $n$-component
column vector. The polynomial solutions $w_\bk$ have inverse Fourier transforms
that are compactly supported. But when the filled band is determined by such a set of polynomial equations, 
we call it a {\em polynomial bundle} \cite{dubail}; this should not be confused with a 
polynomially-generated bundle,
because in general the set of all solutions (together with their translations in position space) may 
not contain a set of Wannier-type functions.

However, if there is a gap in the energy spectrum that separates the filled from the empty states at
each $\bk$, then the vector bundle formed by the filled-band states will be analytic. In Ref.\
\cite{dubail}, it was shown that for a polynomial bundle that is analytic, for each $\bk$ there is a set of
$m$ polynomial sections [solutions to eq.\ (\ref{poly_eq})] that span the fibre of the filled band bundle
in a neighborhood of that $\bk$. Thus for an analytic polynomial bundle there are
compactly-supported Wannier-type functions, and conversely it is easy to see that if there are 
compactly-supported Wannier-type functions, then the vector bundle is analytic, though not necessarily 
a polynomial bundle.

A further strong result, called a no-go theorem, was proved by DR in Ref.\ \cite{dubail}. It says that
for band structures as discussed, if the filled-band bundle is polynomial and analytic, then it is 
topologically
trivial as a complex vector bundle. (A heuristic argument for a version of this statement in two space
dimensions was given in Ref.\ \cite{hast_notes}.) This then implies that if a TNS state of this type is
constructed with a non-trivial vector bundle, then it is non-analytic, and hence any parent Hamiltonian
for it must be gapless.
Here a single-particle {\em parent Hamiltonian} is one for which there is a set
of energy bands in which the eigenstates span the fibre of the ``filled-band'' bundle at each $\bk$, and the
remainder span the empty-band bundle, and further the Hamiltonian has strictly short-range matrix elements.
The latter condition is equivalent to having Laurent polynomial matrix elements in $\bk$ space (for a
Hamiltonian, one cannot reduce the problem to ordinary polynomials, because the Hamiltonian has to be
Hermitian). More generally, the result implies that any short-range single-particle Hamiltonian (with,
say, exponentially-decaying matrix elements) for which the states in the filled and empty (i.e.\ positive
and negative energy) bands span the given non-trivial polynomial filled- and empty-band bundles must 
be gapless.

A modified version of the proof of the no-go theorem also shows \cite{dubail} that if there are
compactly-supported Wannier-type functions for a vector bundle, then the vector bundle is topologically
trivial. This statement is more
general than the no-go theorem for a free-fermion TNS, because for the latter the polynomial sections are
defined by polynomial equations, and this is presumably less general than simply having a set of sections
given.

These statements about triviality should be interpreted with care, because in fact the proofs in 
DR in general establish only that the bundle is {\em stably} topologically trivial, but not 
necessarily topologically trivial; however, for $d\leq 3$, these notions are equivalent to each other 
and to the vanishing of all Chern numbers. We define and explain the notion of stable triviality, 
which is natural in $K$-theory, in Sec.\ \ref{syzygy} below.

\subsection{Flat-band Hamiltonians}

A further area where similar ideas have appeared is flat-band Hamiltonians for non-trivial vector bundles;
here a flat energy band is an energy band in which the energy eigenvalue is independent of $\bk$ over
the whole Brillouin torus. When more than one flat band is present, the case of interest is usually that in
which there are flat bands that all have the same energy eigenvalue (regardless of whether or not the
remaining bands are also flat, with different energy). A particular question that has appeared
\cite{seidel} is whether the Hamiltonian that has the flat band or bands can have strictly-short-range
matrix elements. (This paper appeared earlier than the
published version of Ref.\ \cite{dubail}, but later than the first version and Ref.\ \cite{cirac}; it was
unfortunately not known to us until a late stage in the present work.) We note
immediately the similarity to the parent Hamiltonians discussed in the preceding Subsection. In Ref.\
\cite{seidel}, the authors proved that for such a two-dimensional Hamiltonian with a single flat band, or
a degenerate set of flat bands, the Chern number of the (set of) band(s) must vanish. Further work on
related problems appears in Ref.\ \cite{arovas}.

We can give a short proof of the result of Ref.\ \cite{seidel} as a consequence of the no-go theorem
mentioned in the preceding subsection.
First, we notice that the eigenvalue problem for the flat band (or degenerate flat bands) is given by
\be
{\cal H}_\bk w_\bk=0,
\ee
where $w_\bk$ is again a section of the vector bundle defined by the flat bands, ${\cal H}_\bk$ has Laurent
polynomial entries, and we have set the energy of the states in the flat band to zero by a shift of the
Hamiltonian by a multiple of the identity if necessary. After multiplying by positive powers of $X_\mu$s
(as necessary), these equations have the same form as the polynomial equations
(\ref{poly_eq}). Thus the flat-band vector bundle is polynomial. If we assume that an energy
gap is present above and below the flat band at all $\bk$, then it is also analytic; this assumption
seems to be implicit in the problem at hand. Then it follows from the DR no-go theorem \cite{dubail} that
the vector bundle is stably trivial. This includes the $d=2$ result of Ref.\ \cite{seidel}, and in fact 
goes further, as it applies in all dimensions $d$ of space.

The flat-band problem is less general than the TNSs, because in the former case the polynomial equations
involve the Hamiltonian, not a more general matrix, and the Hamiltonian must be Hermitian at each $\bk$,
unlike the $Z_\bk$ matrix above. We see that the problem of compactly-supported Wannier-type functions
is the most general of all.


\subsection{Symmetry classes, role of algebraic $K$-theory, and results}

The full topological classification of the ten symmetry classes of band structures for non-interacting
particles (or linear wave equations), with a gap in the energy spectrum, on a lattice with translation,
but no other crystallographic,
symmetries was introduced in Refs.\ \cite{srfl,kitaev}. It was connected with topological $K$-theory
by Kitaev \cite{kitaev}; see also Refs.\ \cite{stone,fm}. The basic meaning of each of the ten classes
will be explained later in the paper as we go through the cases.

In the present paper, the goal is to generalize the no-go theorems already mentioned, which were for the
general case of band structures for complex hopping Hamiltonians, or for complex vector bundles, to all
of the ten symmetry classes that occur when the system has translation symmetry but no other
crystallographic symmetries. These will include the paired states, or superconductors, for which particle
number is not conserved; these can always be mapped onto number-conserving single-particle models by
``doubling'' in a well-known way.

If one starts by examining the compactly-supported functions, or polynomial sections, it soon becomes
apparent that an algebraic approach may be fruitful. As explained above, given a finite set of
compactly-supported functions, one can obtain others by applying translations in position space, and
by taking linear combinations of the functions and their translates. (Our analysis always assumes that the
lattice is infinite, though the results nonetheless have applications in finite systems with periodic
boundary conditions.) We restrict the linear combinations
to consist of finitely-many terms, each of which is some translate of one of the (finite) initial set
of compactly supported functions; this condition ensures that the combinations are again compactly
supported. In Fourier ($\bk$) space, translation in position space becomes multiplication by factors
like $X_\mu$ (defined above) to positive or negative integer powers, and so we are taking linear
combinations of the given set of polynomial sections, with coefficients
that are polynomials in $X_\mu$ with complex coefficients in the polynomial (in the simplest case of
complex vector bundles without further symmetries, as in all examples so far). The set of all such
combinations forms what is called a ``module'' over the ring of such polynomials, and the module is said
to be ``generated'' by the initial compactly-supported functions or polynomial sections; this is analogous
to having a vector space that is generated (spanned) by a given finite set of vectors, meaning that
all others are obtained as linear combinations of the latter with complex coefficients. The difference 
is that the ``scalars'' are
now taken in a ring (of polynomials), rather than belonging to the field of complex numbers, while the
vectors are actually vectors of polynomials in $X_\mu$s (and so both the scalars and the vectors can
be viewed as functions of $\bk$), and this makes a significant difference to the structure, which can be
much less trivial than it is for vector spaces with complex scalars.

It turns out that algebraic $K$-theory provides appropriate tools for classifying this structure.
[For experts, we mention that we need only the ``lower'' algebraic $K$-theory of Grothendieck
and Bass, not the ``higher'' theory of Milnor and Quillen.] In
particular, our goal is to classify which of the topological classes of bundles in each symmetry class can
be generated by a set of polynomial sections; this classification is the desired extension of the no-go
theorem to the remaining nine symmetry classes. For this, we need to associate a topological $K$-theory
class (or element of the $K$-group, or values of a complete set of numerical invariants characterizing
such classes or elements), as in Ref.\ \cite{kitaev}, with the bundle generated by the compactly-supported
Wannier-type functions. This information can be obtained from algebraic
$K$-theory groups because of the existence of natural maps from the latter into the topological $K$-group
classification (a similar approach was also used in Ref.\ \cite{seidel}); hence it is useful first to
classify the possible modules using algebraic $K$-theory, for each of the ten symmetry classes, before
checking that the maps to the topological theory work properly.

The final results can be characterized as follows. One does not quite have a no-go theorem, saying that
no non-(stably)-trivial band structure can be obtained, but instead there is a very limited set of 
possibilities.
To describe this, we point out that (in some of the symmetry classes, though not for the complex vector
bundles as in DR), there can be a non-trivial
``winding number'' of the vector bundle (or of some aspect of it) as a function of $\bk$ along a closed
path in the Brillouin torus that winds around it, for example a path along one of the coordinate axes.
Such winding can occur in that symmetry class in any dimension, and in the case of $d=1$ dimension
it provides the only possible non-trivial invariant of the bundles, taking values in some cases in the
integers $\bf Z$,
and in other cases in ${\bf Z}/2$ (read as ``$\bf Z$ mod $2$'', i.e.\ the group with two elements); in
five of the ten symmetry classes, the invariant can only be zero. In higher dimensions for the symmetry
classes, the same values as in one dimension for the same symmetry class for these winding-number
invariants can occur independently for each of the $d$ directions of $\bk$-space, giving groups consisting
of $d$-tuples of elements of either $\bf Z$ or ${\bf Z}/2$, or else the trivial group.
In the topological classification, there are topological classes distinguished by non-zero values of
other invariants (such as a Chern number) as well as by the winding numbers. But the result of the
algebraic analysis for the bundles obtained from compactly-supported Wannier-type functions is that the
only non-trivial instances are those with non-trivial winding numbers as just described, and which are
trivial in all other ways, except in some cases for a global invariant coming from zero dimensions. 
In zero and one dimension, these cover all topological classes, but in more than
one dimension these instances are particular examples of ``weak'' topological insulators
or superconductors, in which the only non-trivial topology essentially arises from what is possible in
one dimension in that symmetry class, applied to each of the $d$ directions of $\bk$-space, together with 
what is possible in zero dimensions. For $d>1$,
none of the ``strong'' topological insulators or superconductors can be obtained from compactly-supported
Wannier-type functions or as free-fermion TNSs. For the reader's convenience, these results are spelled
out in full detail, without the technical derivation,
in section \ref{disc}, where a Table also appears that makes comparison with the general case in $0$,
$1$, and $2$ dimensions.

The plan of the remainder of the paper is as follows: section \ref{math} gives some of the mathematical
background used. Sections \ref{wdclass}, \ref{chiral}, and \ref{sec:az} give the detailed analysis, first
for the three
classic Wigner-Dyson classes (section \ref{wdclass}), then for the three chiral symmetry classes (section
\ref{chiral}), and finally for the remaining Altland-Zirnbauer classes (section \ref{sec:az}). Each of
these sections consists of two parts, the first describing the relevant features of the symmetry classes,
the second the algebraic $K$-theory analysis of them (including the mapping into
topological $K$-theory). Much of the structure of the arguments in the later sections is the same as that
in section \ref{wdclass}, which should be read carefully. Section \ref{disc}, as mentioned, includes a
summary of the precise results,
and also some discussion of the underlying reasons for the results, and a conjectured extension to
interacting systems. The final section is a Conclusion.

\section{Mathematical background}
\label{math}

\subsection{Algebraic background and definitions}
\label{background}

This section mostly provides background, but can be skimmed and referred back to later. For this general
background, see also Refs.\ \cite{reid,jacob,anf,passman}.

We recall that a {\em ring} is a set of elements that forms an Abelian group under addition (with $0$ as
the identity) and has an associative multiplication operation that distributes over addition; we denote
a generic ring by $R$. All our rings
have the multiplicative identity element, written $1$. A ring is called commutative if multiplication
is commutative. The elements in a ring $R$ that have a multiplicative inverse in $R$ are called the units,
and form a multiplicative group denoted $R^\times$. A division ring is a ring in which every non-zero
element is a unit; a commutative division ring is a field. A ring $R$ is an algebra over a field $\bf F$ if
the center of $R$ (the set of elements that commute with all elements) contains a
copy of the field $\bf F$ (with the multiplicative identity identified with that in the
field); thus an algebra is a vector space over $\bf F$ with an associative and distributive 
multiplication (with an identity)
defined on the vectors. Our most basic examples of rings are the integers $\bf Z$, real numbers
$\bf R$, complex numbers $\bf C$, and quaternions $\bf H$. The latter are defined as linear combinations,
using real coefficients, of the identity $1$ and elements $\hati$, $\hatj$, $\hatk$ (not to be confused
with vectors $\bk$) subject to the relations
\be
{\hati}^2={\hatj}^2={\hatk}^2={\hati}{\hatj}{\hatk}=-1.
\label{quatrels}
\ee
$\bf R$ and $\bf C$ are of course fields, and $\bf H$ is a non-commutative division ring. $\bf R$, $\bf C$,
and $\bf H$ are also algebras over the field $\bf R$, and are the only finite-dimensional associative 
division algebras over $\bf R$.

We will also use some polynomial rings. The simplest are of the form $R[X_1,\ldots,X_d]$, with
$d\geq0$ indeterminates $X_\mu$, $\mu=1$, \ldots, $d$, and consist of polynomials in the $X_\mu$s
with coefficients in a ring $R$; addition and multiplication are defined in an obvious way. We will also
use extensively the rings of Laurent polynomials, denoted
$R[X_1,X_1^{-1},X_2,X_2^{-1},\ldots,X_d,X_d^{-1}]$,
which are polynomials with both positive and negative powers, but with the exponent of each $X_\mu$ in
a polynomial bounded both above and below. More generically these two types of polynomial rings will
be written $R[X_\mu]$ and $R[X_\mu^{\pm 1}]$. For the following Laurent polynomial rings we will use
notation
\bea
R_1&=&{\bf C}[X_1,X_1^{-1},\ldots,X_d,X_d^{-1}],\nonumber\\
R_2&=&{\bf R}[X_1,X_1^{-1},\ldots,X_d,X_d^{-1}],\nonumber\\
R_3&=&{\bf H}[X_1,X_1^{-1},\ldots,X_d,X_d^{-1}].
\eea
or $R_i^{(d)}$ ($i=1$, $2$, $3$) when we wish to specify the space dimension $d$. The last of these
rings $R_3$ is not commutative. Each of them contains an image of the underlying ring $R$, consisting of
the constant polynomials (with no $X_\mu$ appearing in the expression). Further, the units (invertible
elements) in the polynomial rings $R[X_\mu]$ with $R = {\bf R}$, $\bf C$, or $\bf H$ are precisely
the non-zero constants, while those in the Laurent polynomial rings $R_i$ are the monomials, of the form
\be
c \prod_\mu X_\mu^{m_\mu}
\label{monom}
\ee
where $c$ is a nonzero constant and $m_\mu$ are integers, as may easily be checked. All polynomial rings
(Laurent or not) over a division ring are both right and left Noetherian (see Refs.\
\cite{reid,jacob,passman} for the definition and the result);
the distinction between right and left lapses for commutative rings. $R_1$
and $R_2$ are unique factorization domains (as is $\bf Z$), that is any element [a (Laurent) polynomial]
can be factored into prime or irreducible polynomials, and the prime factorization is unique up to
permutation of the factors and multiplication of factors by units. [This is well-known for the
ordinary polynomial rings over a field \cite{reid}, and for Laurent polynomials follows by shifting
exponents (by multiplying by units) until all exponents of all $X_\mu$ are non-negative.]

We will use {\em modules} over various rings. A module $M$ over $R$ (an $R$-module) is a set of elements
that form an Abelian
group (written additively), with an action of the ring $R$ taking any element of the module to some other
element, written as multiplication: if $m\in M$, $r\in R$, then $m\rightarrow mr$ is the map,
with $(m_1+m_2)r = m_1 r+m_2 r$ and $(mr_1)r_2=m(r_1r_2)$. Notice that we write the element of
the ring acting from the right, so all our modules are right modules unless otherwise stated;
for commutative
rings $R$, a right module can be viewed as a left module (or {\it vice versa}), but for noncommutative
rings, right and left modules are distinct. (Many properties of a ring are module-theoretic in character,
and so, in the non-commutative case, are defined for right or left action, as for
``Noetherian'' which was already mentioned. When the term ``right'' appears before the name of a property,
it means that there is a parallel definition for the ``left'' version.) The ring $R$ is both a left and
right module over itself (i.e.\ a bimodule). A homomorphism from one module to another, both over the
same ring, is a ``linear'' map that commutes both with addition and with the action of the
ring on the modules. An isomorphism is a homomorphism that has an inverse homomorphism; we write $A\cong
B$ when an isomorphism exists. Usually our modules $M$
will be {\em finitely generated} (f.g.), that is, there is a {\em finite} set of generators in $M$ such
that any element can be expressed as a linear combination of the generators, with coefficients in $R$.
A submodule of a module is any subset that also forms a module. A direct sum, written $M_1 \oplus M_2$
of modules over $R$ is really a module consisting of all pairs $(m_1,m_2)$ with $m_1\in M_1$, $m_2\in
M_2$, with addition of pairs, and multiplication of a pair by an element of the ring, defined
componentwise. It will
be common to say that some module ``is'' a direct sum if it is isomorphic as a module to a direct sum
of modules; in this case the module is ``decomposable'' as a direct sum. It is important that in general
when a module has a submodule, the module is not necessarily a direct sum (unlike for representations
of finite groups, for example).

Different types of modules will enter this work. Some are free modules, which can be generated by $n$
generators that are linearly independent over the ring. (We will usually only require f.g.\
free modules.) Using the generators, a f.g.\ free module, say $F$, can be
represented faithfully as the set of all column vectors with entries in $R$, and so is isomorphic 
to $R^n$ (the iterated
direct sum of $n$ copies of $R$) for some $n$.
Such a free module is said to have rank $n$. If the ring is right Noetherian, then any submodule of a
finite-rank free module is finitely generated \cite{reid,jacob,anf,passman}. As examples, we mention 
that modules
over a division ring are always free, and can be termed ``vector spaces'' over the division ring.
This includes the case of non-commutative division rings such as the quaternions $\bf H$ \cite{artin}, 
as well as the fields $\bf R$ and $\bf C$.

An important tool is the idea of an exact sequence. If $A$, $B$ $C$ are modules over $R$, then a pair
of maps (homomorphisms) $\phi_1:A\longrightarrow B$ and $\phi_2:B\longrightarrow C$ form an exact sequence
\be
A\stackrel{\phi_1}\longrightarrow B\stackrel{\phi_2}\longrightarrow C
\ee
if and only if the image ${\rm im}\, \phi_1$ of $A$ under $\phi_1$ is precisely the kernel $\ker \phi_2$
of $\phi_2$ (both the kernel
and the image of a homomorphism are modules). Thus not only do they compose to give the zero map
$\phi_2\circ\phi_1=0$ from $A$ to $C$, but the first map is a surjection onto the
kernel of the second. When more maps are present, as in
\be
A_1\stackrel{\phi_1}\longrightarrow A_2\stackrel{\phi_2}\longrightarrow \cdots
\stackrel{\phi_{n-1}}\longrightarrow A_n,
\ee
then the statement that the sequence is exact means that exactness holds at each term at which there is
both a map in and a map out, as for $A_2$ through $A_{n-1}$ here.
In particular, a ``short exact sequence''
\be
0\longrightarrow A\stackrel{\phi_1}\longrightarrow B\stackrel{\phi_2}\longrightarrow C\longrightarrow 0
\label{short_exact}
\ee
means that $C\cong B/A$ as a module, or strictly $C\cong B/{\rm im}\,\phi_1$. In this case, one can say
that $A$ is (isomorphic to) $\ker \phi_2$, while $C$ is (isomorphic to) the {\em cokernel} of $\phi_1$,
that is $B/{\rm im}\,\phi_1$.

Another important class of modules are the projective modules \cite{jacob,anf,passman} (they must not be
confused with projective representations, which are entirely different). They can be defined in several
ways. One
way is as a module that is isomorphic to a summand in a free module. Thus $P$ is projective if and only
if there is a $P'$ such that $P\oplus P' \cong F$ for some free module $F$ (it follows that
$P'$ is also projective); for $P$ finitely generated, $P'$ and $F$ can be taken to be
finitely generated. Clearly, any free module is projective. Another way to define a projective module is by
saying that any short exact sequence ending in a projective module {\em splits}, that is, $P$ is 
projective if and only if for any short exact sequence ending in $P$,
\be
0\longrightarrow A\longrightarrow B\longrightarrow P\longrightarrow 0,
\ee
we have $B\cong A\oplus P$.

We will sometimes need the general notion of a tensor product over a ring that may be non-commutative.
If $M_1$ is a right module and $M_2$ is a {\em left} module over $R$, then the tensor product
$M_1\otimes_R M_2$ is generated (over $\bf Z$, in the basic case) by the set of pairs $(m_1,m_2)$ of
elements $m_1\in M_1$, $m_2\in M_2$, modulo
relations that make it bilinear in $m_1$ and $m_2$ under addition, and also such that elements of $R$ can
be moved between the factors: $(m_1r)\otimes m_2=m_1\otimes rm_2$ for all $m_1$, $m_2$, $r\in R$. The
tensor product is not always a module over $R$, though it is always a module over $\bf Z$. But for
a bimodule, one does get a module. For example, as $R$ is a right-left $R$-bimodule, for any right
$R$-module $M$, $M\otimes_R R\cong M$ as a right module.

Sometimes it is desired to relate modules for one ring to those of another, when the rings can be
related. Given a homomorphism from one ring $R$ to another $S$, say 
$\widehat{\varphi}:R\longrightarrow S$, modules over $R$ and $S$ can be related. Suppose for simplicity 
(as for the case we will use) that $R$ is a subring of $S$, so $\widehat{\varphi}$ is an inclusion. 
One way to relate the respective modules is via the {\em pullback} or forgetful map: in view of the 
inclusion, a module $M$ over $S$ is automatically a module over $R$. Formally this can be expressed 
using the tensor product, because $S$ can be viewed as a left $S$-module and as a right $R$-module, 
so $M\otimes_S S$ (which is isomorphic to $M$ as an $S$-module) is a right $R$ module. On the other 
hand, there is also the {\em change-of-rings} map. Given a right $R$-module $N$, and using $S$ viewed 
as a left $R$-module and right $S$-module, $N\otimes_R S$ produces a right $S$-module, which is likely 
to be larger than $N$. For example, in the context of representation theory of groups, one studies 
modules over the group algebra. For a subgroup $H$ of a group $G$, there is a corresponding inclusion 
of group algebras, and the pullback and change-of-rings maps are known as restriction and induction, 
respectively. It will be helpful to realize that when one has a set of generators for an $R$-module, 
it can be proved that the change-of-rings map produces a corresponding set of generators for the 
resulting $S$-module; in particular, the latter set of generators has the same cardinality as the former.

Finally, we should mention that the collection of all modules over a ring $R$, together with the
homomorphisms between them, form a {\em category} \cite{jacob,anf,passman}. The modules are the objects, 
and the
homomorphisms are the maps (or arrows, or morphisms) of the category; a morphism into a module can be 
composed with a morphism out to yield another morphism, and there is a unique identity morphism from 
each object to itself. (The subcollection of all f.g.\ modules together with all homomorphisms between 
them forms a ``full'' subcategory, as does the collection of all f.g.\ projective modules likewise). 
Maps between categories are called {\em functors}; one has to specify an image under the functor for 
each object and for
each morphism, with the condition that the functor respects composition of morphisms and the identity
morphisms. The pullback and change-of-rings maps mentioned above in fact define functors between the
categories of modules of the two rings.

\subsection{Vector bundles as modules}
\label{bunmod}

We have already discussed the notion of a vector bundle. Here we want to relate vector bundles to
modules over a ring, and so make contact with the algebraic approach. First, given a (finite-rank)
complex vector bundle over a base space $B$ (such as a sub-bundle of a rank-$n$ trivial vector bundle,
as for the vector bundle associated to some bands in a tight-binding model; then $B$ is the Brillouin
torus), we can consider the space of all its sections. It is clear that these form an
(infinite-dimensional) vector space over $\bf C$, and that the vector bundle
can be recovered from its space of sections. (This space is not a Hilbert space, but can be completed
to obtain a ``single-particle'' Hilbert space consisting of ``states'' in the bundle, expressed as
vector-valued functions on $B$, by using a non-degenerate inner product on each fibre, and integration
over $B$ with some measure, to obtain the inner product on an $L^2$ space formed from the
bundle. For the case when $B$ is the
Brillouin torus, this Hilbert space is equivalent to that of single-particle states in the original
lattice. This is a good place to point out that, except for passing inessential references to
orthonormality or to unitary matrices on the fibre, we make no use of these inner products on the fibre
and on the bundle in the arguments in this paper.) If we introduce the ring $C_{\bf C}(B)$ of
continuous complex functions on $B$, then because we can take linear combinations of sections using
elements of $C_{\bf C}(B)$ as coefficients, the space of sections is in fact a module over $C_{\bf C}(B)$.

Further,
Swan's theorem \cite{swan} says that if $B$ is a compact Hausdorff space, then a module
over $C_{\bf C}(B)$ is isomorphic (as a module over $C_{\bf C}(B)$) to the space of sections of a
vector bundle if and only if the module is finitely generated and projective. Thus, when $B$ is compact
and Hausdorff, the f.g.\ projective $C_{\bf C}(B)$-modules are precisely the spaces of sections of
vector bundles. (It is clear that a free $C_{\bf C}(B)$-module corresponds to a trivial vector bundle 
over $B$.) Being projective means that for any vector bundle (such
as a filled-band bundle in our case), there is another vector bundle such that the direct sum (i.e.\
the direct sum of the fibres at each $\bk$, also known as the Whitney sum) of the two is a trivial
vector bundle (indeed, in our basic example, we also have an empty-band bundle, and these are two subspaces
of the fibre at each point in $\bk$ space, so the direct sum is the trivial vector bundle of the
tight-binding model).
The condition that the rank of the vector bundle is constant on $B$ (which we can assume is connected)
is necessary for this to be valid. Being finitely generated means that there is a {\em finite} set of
sections that generate the module, that is such that combinations of them (with continuous complex
function coefficients) span the space of sections, or in particular span the fibre at each point of $B$.
The assumptions that
$B$ is compact and Hausdorff ensure that this is true for a vector bundle. We note that similar statements
can be made for types of vector bundles other than complex ones, such as the ones we will encounter later.

Our definition of a set of Wannier-type functions (after Fourier transform) was a set of sections that 
span the fibre at all points $\bk$ in $B=T^d$, the $d$-dimensional (Brillouin) torus $T^d$, and hence 
which generate the space of sections as a
module over the ring $C_{\bf C}(B)$. Hence part of Swan's theorem guarantees that the set can be assumed 
to be finite. Notice that the mathematical argument (and our definition) only required sections to 
be continuous, and that if stronger smoothness conditions are placed on the sections (so that in 
position space they decay rapidly), this might change the result; whether a finite set obeying such 
conditions exists in general is outside the scope of this paper. We will be discussing finite sets of 
Wannier-type functions that are analytic, indeed polynomial, sections.

When we turn to compactly-supported packets within some set of bands, in $\bk$ space each one corresponds
to a section of the vector bundle, and in our standard basis derived from the tight-binding model these are
vectors with Laurent polynomial entries. If we have a set of compactly-supported Wannier-type functions,
then we have a set of polynomial sections that generate the projective module of all continuous sections.
We will call this a set of polynomial generators for the module (or by abuse of language, the vector
bundle), and say that a vector bundle that is a subbundle of a trivial vector bundle and has a set
of generators that are polynomial sections is a {\em polynomially-generated} module (over $C_{\bf C}(B)$)
or vector bundle. This terminology is briefer than saying that the vector bundle admits a set of
compactly-supported Wannier-type functions; we note that we already used the term polynomial bundle for
a {\em different} notion.

The ring $R_1$ of complex Laurent polynomials can be related to $C_{\bf C}(B)$ when $B=T^d$ by 
evaluating each $X_\mu$ as a complex number with $|X_\mu|=1$.
A Laurent polynomial then becomes a continuous function on the torus [these functions are dense in
the sup-norm topology on $C_{\bf C}(T^d)$, but we make no use of this fact]. Hence we have a homomorphism
of rings $R_1\to C_{\bf C}(T^d)$, which is injective, so $R_1$ is a subring of $C_{\bf C}(T^d)$. A module
over $R_1$ consisting of some set
of $n$-component vectors with polynomial entries (i.e.\ a submodule of a free module) then produces a
module over $C_{\bf C}(T^d)$ simply by combining the polynomial vectors (sections) using arbitrary
continuous complex-function coefficients. This is an instance of the change-of-rings functor
corresponding to the inclusion $R_1\subseteq C_{\bf C}(T^d)$. (This functor was already used in Ref.\
\cite{seidel} and in a less formal way in DR \cite{dubail}.) This functor always maps a projective module
to a projective module (a vector bundle). However if the $R_1$-module is not projective, the resulting
$C_{\bf C}(T^d)$-module may not be projective, and so may not correspond to a vector bundle. If there is a 
finite set of generators (consisting of $n$-component vectors with polynomial entries) for the 
$R_1$-module, then after change of rings those generators are viewed as $n$-component vectors with 
polynomial functions of $\bk$ as entries, and generate a $C_{\bf C}(T^d)$-module, which is a submodule of 
the free module $C_{\bf C}(T^d)^n$. Thus, our condition that the compactly-supported
Wannier-type functions span the fibre (with constant rank) at each $\bk$ ensures (by construction) that 
the change-of-rings map produces a polynomially-generated bundle, or in other words, that the resulting 
module over $C_{\bf C}(T^d)$ is in fact projective. This condition is weaker than the condition of being 
a projective $R_1$-module. In order to obtain results about vector bundles, the use of this condition 
will be crucial to our treatment.


\section{Wigner-Dyson classes A, AI, AII: $K_0(R)$}
\label{wdclass}

Now we begin to describe the extension of the no-go theorem to other symmetry classes. The simplest cases
are the classic Wigner-Dyson symmetry classes, which (like the others) originated in the context of
random matrix theory. These are known as the unitary, orthogonal, and symplectic ensembles, or as
symmetry classes A, AI, and AII. The unitary class A was already covered \cite{dubail}, but we will
include some review of that case here. In these classes the basic issues involve vector bundles of certain
types; this part of the discussion is also relevant for the other classes later.

\subsection{Cases AI, AII}
\label{aiaii}

We will begin with general descriptions of these symmetry classes in the context of translation-invariant
single-particle Hamiltonians. The use of a Hamiltonian here is solely to motivate the symmetry structure;
the arguments in the proof refer only to the module of sections of the filled-band bundle, not to a
Hamiltonian.

As mentioned already, for class A the single-particle Hamiltonian $\cal H$ is allowed to be complex,
and in $\bk$ space is a Hermitian $n\times n$ matrix for each $\bk$, continuous (or even smoother) in
$\bk$. For the orthogonal class AI, we simply require the Hamiltonian $\cal H$ in position space to be
real; we can think of this as a statement of time-reversal symmetry, implemented by
an antiunitary operator $\widehat{T}$ with $\widehat{T}^2=+1$, and the time reversal operation
$\widehat T$ reduces to complex conjugation $\widehat{T}=K$; thus $\widehat{T} {\cal H}
\widehat{T}^{-1}={\cal H}$ means $\cal H$ is real. (Strictly, this statement is basis
dependent, and one should say that there exists a basis in which the Hamiltonian is real; by saying the
Hamiltonian is real we have chosen such a basis once and for all. In fact, we are assuming that it is real
in a basis of the form natural for a tight-binding model, as already defined.) In $\bk$ space,
the Hamiltonian splits into blocks ${\cal H}_\bk$ labeled by $\bk$, which are Hermitian. Now the
matrix elements in $\bk$ space are the Fourier transforms of corresponding ``hopping'' functions $f(\bx)$
(for a displacement by $\bx$) in position space, which are real. The Fourier transform $f_\bk$ of such
a function obeys
\be
f_{-\bk}=\overline{f_\bk},
\ee
where $\bar{}$ is complex conjugation.
Atiyah \cite{atiyah} calls such a function (of $\bk$) Real (with a capital R) instead of real. In fact,
he defines a Real space $B$ (such as our Brillouin torus) to be one with an involution that sends a
point $x\in B$ to a ``conjugate point'' $\bar{x}\in B$ (where the bar does {\em not} mean complex
conjugate), with $\bar{\bar{x}}=x$.
In our case, the map is $\bk\longrightarrow -\bk$, which is well defined modulo
reciprocal lattice vectors. We may now describe our time-reversal--invariant Hamiltonian ${\cal H}_\bk$
by saying that its matrix elements are Real; it obeys
\be
\overline{{\cal H}_\bk}={\cal H}_{-\bk},
\ee
where $\bar{}$ on a complex matrix stands for complex conjugation of each matrix element.

Because the Hamiltonian in position space is real and symmetric, its eigenvectors can be chosen to
be real. In $\bk$ space, these become Real vectors, and we can speak of Real sections of the vector bundle;
the
inverse transform of a Real section is a real wavepacket in position space. In this case,
the filled-band bundle has a Real structure \cite{atiyah}, that is a map of the total space that
sends vectors in the fibre at $x$ to ones in the fibre at $\bar{x}$, which is antilinear (like
complex conjugation) on each fibre and squares to the identity. Hence the vector bundle formed by the
filled-band
states is a Real vector bundle. Without loss of generality, it can be studied as a module (over the ring of
continuous Real functions) consisting of Real sections only. Clearly there are sections of the (Real)
vector bundle that are not Real, however, any such section can be decomposed as a sum of two Real sections
(using the Real and ``Imaginary'' parts), so no information is lost. This is similar to studying a
complex vector space with a real structure (i.e.\ the operation of complex conjugation) in terms of
real vectors only. Note that for the eigenvectors of the Hamiltonian, the Real symmetry implies that
if $w_\bk$ is an eigenvector of ${\cal H}_\bk$ with energy eigenvalue $E_\bk$, then $\overline{w_\bk}$ is
an eigenvector of ${\cal H}_{-\bk}$ with the same energy eigenvalue $E_{-\bk}=E_\bk$, and we can choose
phases and identify the eigenvectors as $w_{-\bk}=\overline{w_\bk}$. When this holds for all $\bk$,
these are vectors with Real entries.

When we turn to compactly-supported functions and polynomial sections, we must consider Real polynomials.
These are polynomials in the $X_\mu$s, and should be Real functions. But the conjugate of
$X_\mu=e^{ik_\mu}$ is, for real $\bk$, just $X_\mu$ evaluated at $-k_\mu$. So the involution on $T^d$
leaves $X_\mu$ invariant. Then Real polynomials are simply polynomials in the $X_\mu$s with real
coefficients; they form the ring $R_2$ already defined.

Now we turn to the symplectic class AII. In this case we think of spin-1/2 particles, and there is
time-reversal symmetry acting in the Kramers mode, with $\widehat{T}^2=-I$ in the single-particle Hilbert
space. For a general one-particle Hamiltonian $\cal H$ acting in a finite-dimensional Hilbert space of
orbitals for either spin, we define $\widehat{T}$ (with conventional choice of basis) to be
\be
\widehat{T}=KU,
\ee
where $K$ is complex conjugation, and $U$ is unitary, with
\be
U=i\sigma_y\otimes I
\ee
(where the second factor is the identity on the space of orbitals, the first acts in the spin space,
and
\be
\sigma_x=\left( \begin{array}{cc}0&1\\1&0\end{array}\right),\;
\sigma_y=\left( \begin{array}{cc}0&-i\\i&0\end{array}\right),\;
\sigma_z=\left( \begin{array}{cc}1&0\\0&-1\end{array}\right)
\ee
are the usual Pauli matrices). We write matrices like $\sigma_\alpha\otimes I$ generically
as $\Sigma_\alpha$
($\alpha=x$, $y$, or $z$) for a tensor product space of this form for any dimension of the second factor,
and also $i\Sigma_y$ as $J$, so $\widehat{T}=KJ$.

Time-reversal symmetry means that
\be
\widehat{T}{\cal H}\widehat{T}^{-1}={\cal H}.
\ee
Because of the structure of $J$, this can be reduced to a similar condition for the $2\times2$ blocks of
$\cal H$ in the first factor in the tensor product. The time-reversal--invariant $2\times 2$ blocks can
be expressed as linear combinations, with real coefficients, of the $2\times 2$ matrices \cite{jacob}
\be
1,\;{\hati} = i\sigma_z,\;{\hatj}=i\sigma_y,\;{\hatk}=i\sigma_x
\ee
(note the ordering of the indices). These obey the relations in eq.\ (\ref{quatrels}) of the generators
of the quaternions, so we used the same symbols. Thus the matrix representing a quaternion
$q=a+b{\hati}+c{\hatj}
+d{\hatk}$, where $a$, $b$, $c$, $d$ are real, has the form
\be
q=\left( \begin{array}{cc}a+ib&c+id\\-c+id&a-ib\end{array}\right),
\label{quater}
\ee
and there is a natural injective map of the real numbers $\bf R$ into $\bf H$ that maps the real number
to $a$. Notice that the determinant of the matrix is $a^2+b^2+c^2+d^2=|q|^2$, which defines the norm 
$|q|\geq0$ of the quaternion, a real number. One can define a ``conjugation'' operation on the
quaternions, $q\to \overline{q}$ ($q\in {\bf H}$), which is an isomorphism that reverses the order in
a product of quaternions, by
\be
\overline{\hati}=-{\hati},\;\overline{\hatj}=-{\hatj},\;\overline{\hatk}=-{\hatk},\;
\ee
while $\overline{1}=1$. (When quaternions are expressed as $2\times 2$ matrices, this is
the usual adjoint. We hope that no confusion will arise from the use of the bar $\bar{}$ to represent
both complex and quaternionic conjugation; which is meant should be clear from the the context,
specifically whether complex or quaternionic coefficients are in use.) The norm-square of $q$ is equal
to $q\overline{q}=\overline{q}q=|q|^2$ as a quaternion or as a $2\times 2$ matrix (i.e.\ a non-negative
multiple of the identity). The (right or left) inverse of a quaternion $q$ can be expressed as 
$q^{-1}=\overline{q}/|q|^2$. Then a
time-reversal invariant matrix such as $\cal H$ (Hermitian or not) can be expressed as a matrix of
quaternions. A matrix of quaternions $\cal H$ is Hermitian when viewed as a
complex matrix if and only if it is Hermitian as a matrix of quaternions, where the adjoint
$A\to A^\dagger$ (for a matrix $A$) is defined as
the corresponding conjugate of the transpose of the matrix, and has the usual property
$(AB)^\dagger=B^\dagger A^\dagger$, in either point of view.

In addition, time-reversal applied to an eigenvector (viewed as a column vector of complex numbers)
of $\cal H$, say ${\cal H}\psi=E\psi$, implies that $\widehat{T}\psi$ is also an eigenvector with the
same energy eigenvalue, and not equal to $\psi$. For any vector $\psi$, it and $-\widehat{T}\psi$ can be
assembled into a matrix $v$ with two columns. The two columns are exchanged by time reversal (with a
minus sign in one place, so that $\widehat{T}^2=-1$), as we defined
its action so far. If we define time reversal to act on the matrix (as on any matrix) by
\be
v\to \widehat{T} v \widehat{T}^{-1},
\ee
(using the appropriate size of $J$ in each place) then
$v=\widehat{T} v \widehat{T}^{-1}$, and we can view $v$ as a column vector of quaternions. This shows that
in the symplectic
ensemble, or symmetry class AII, we are in effect dealing with quaternionic vector spaces; using
a basis, maps such as the Hamiltonian act as matrices from the left, as mentioned before, while the
scalar multiplication by
a quaternion is from the right. This relation is mentioned briefly by Atiyah \cite{atiyahbook}, page 33.

When we turn to band structure, we have similar properties for ${\cal H}_\bk$, however again complex
conjugation sends $\bk\to-\bk$. Then the relation is
\be
\widehat{T}{\cal H}_\bk \widehat{T}^{-1}={\cal H}_{-\bk},
\ee
and a similar argument shows that ${\cal H}_\bk$ can be viewed as a matrix with entries that are linear
combinations of $1$, $\hati$, $\hatj$, $\hatk$ with coefficients that are Real functions of $\bk$,
rather than real. It seems reasonable to use the term ``Quaternionic function'' (with a capital $Q$) for
$2\times2$ complex matrix functions of $\bk$ obeying
\be
\widehat{T}f_\bk \widehat{T}^{-1}=f_{-\bk},
\ee
and call the total space a Quaternionic vector bundle (over $T^d$), by analogy with the Real ones.
In addition we have ${\cal H}_\bk^\dagger = {\cal H}_\bk$. Then we can assemble the eigenvectors
$w_\bk$ and $-\widehat{T} w_{-\bk}$ into a $2n\times 2$ complex matrix. Doing so for all $\bk$ gives
a vector with entries that are Quaternionic functions (on which translations in position
space still act as multiplication by powers of the $X_\mu$s), parallel to the Real functions for the AI
case.

If we now consider a Wannier function, then the assumption that time-reversal holds for the
filled-band bundle in question means that the function and its time-reversed partner must both be sections
of the vector bundle. If $w_\bk$ is the transform of a Wannier function, $-\widehat{T}w_{-\bk}$
is (minus) the transform of its time-reverse, and we can form Quaternionic sections (i.e.\ vector
functions of $\bk$, with Quaternionic entries) from these as noted just now. (Then $g_\bk$
also has Quaternionic entries, and $v_\bk=g_\bk u_\bk$ holds for this pair of Wannier functions as
vectors with Quaternionic entries, with notation as in Section \ref{sec:nogo}.) For functions that are also
compactly-supported, we obtain polynomial
Quaternionic sections, that generate a module over ${\bf H}[X_\mu^\pm]=R_3$ or possibly its subring
${\bf H}[X_\mu]$, completely parallel to the complex and Real cases.

It may be helpful here to be explicit about the meaning of a trivial vector bundle for orthogonal and
symplectic
classes. A Real (Quaternionic) vector bundle of rank $m$ over $B=T^d$ is trivial if it has a set of $m$ Real
(Quaternionic) sections that are linearly independent at all $\bk$. (Here linear independence is over $\bf
C$, and so for the Quaternionic case involves $2m$ complex vectors, for $\bk$ such that $\bk\not\equiv
-\bk$; however it reduces to, or can be viewed as, linear independence over $\bf R$ or $\bf H$ when
$\bk\equiv -\bk$.) In all cases,
triviality of a vector bundle of rank $m$ can be viewed as the vector bundle being isomorphic to a product
of $B$ and
${\bf C}^m$ (${\bf C}^{2m}$ in the Quaternionic case), with the obvious action of $\widehat{T}$ on the
vector bundle in the cases of classes AI and AII. The tight-binding model itself has precisely this product
form, with $m$ replaced by $n$.

When $B$ is a Real space, we will use notation $C_R(B)$ [$C_Q(B)$] for the rings of continuous Real
[Quaternionic] functions on $B$, by analogy with the ring $C_{\bf C}(B)$ for continuous complex functions
(in which case the Real structure can be forgotten). Of course, our main interest is in $B=T^d$, the
Brillouin torus. For that case, there are natural embeddings (injective ring homomorphisms) of $R_2^{(d)}$
[$R_3^{(d)}$] into $C_R(B)$ [$C_Q(B)$], similar to that for complex Laurent polynomials and functions
on $T^d$. We will write $C_i$ or $C_i^{(d)}$, $i=1$, $2$, $3$, for the rings $C_{\bf C}(T^d)$, $C_R(T^d)$,
and $C_Q(T^d)$, respectively, so that $R_i^{(d)}\subseteq C_i^{(d)}$ for each $i$ and $d$. A bundle in one 
of the three classes gives rise to its space of sections, which is a module over the corresponding $C_i$; 
for the trivial bundles just discussed, this module is $C_i^m$.

It is now fairly straightforward to extend the proof of the DR no-go theorem \cite{dubail} to the symmetry
classes AI
and AII. First we note that, as in class A \cite{dubail}, it is in fact sufficient to consider polynomial
sections consisting of column vectors with entries in the polynomial rings $R[X_\mu]$ for $R={\bf R}$,
${\bf C}$, ${\bf H}$, with no negative powers; these polynomial rings are more accessible than the
Laurent polynomial analogs. The syzygy theorem holds for polynomial rings with coefficients in any field
\cite{eisen1}, so for ${\bf R}[X_\mu]$ the proof in DR goes over essentially unchanged. The polynomials over
the quaternions form a non-commutative ring, but again a version of the syzygy theorem holds
\cite{passman}.
We discuss these facts in more depth in the section immediately following; modern treatments of them,
especially for non-commutative rings, invariably enter into some $K$-theory.


\subsection{Syzygy theorem and relation with $K_0(R)$}
\label{syzygy}

We will now give some discussion of the syzygy theorem and of its relation to the algebraic $K$-theory
group $K_0(R)$, and give a more conceptual account of the proof of the no-go theorem. In brief outline, 
given the module over the polynomial ring generated by the compactly-supported Wannier-type functions, 
the proof consists of two parts: the syzygy theorem establishes that there is a finite-length free 
resolution of the module over the polynomials, and then the change of rings to the ring of continuous 
functions produces a corresponding free resolution of the vector bundle, from which stable triviality of 
the vector bundle follows (terms used here are defined below). We will explain how the argument and 
result are interpreted in $K$-theory. We repeat that parts
of this discussion are crucial for the cases of other symmetry classes as well. For a nice introduction
to algebraic $K$-theory, see Rosenberg's book \cite{rosenberg}; Refs.\ \cite{milnorK,karoubi,weibelK} are
also useful.

First, we introduce resolutions and the length of a resolution, all for modules over some given ring $R$.
A (possibly infinite) exact sequence
\be
\cdots\longrightarrow P_2\longrightarrow P_1 \longrightarrow P_0\longrightarrow M\rightarrow 0
\ee
is a {\em projective resolution} of a module $M$ if $P_i$ is projective for $i=0$, $1$, \ldots. It is a
{\em free resolution} if each $P_i$ is a free module $F_i$. If a projective resolution terminates at the
left with a zero, say (the labels $\phi_i$ on the maps are for future reference)
\be
0\longrightarrow P_\ell\stackrel{\phi_\ell} \longrightarrow F_{\ell-1}
\stackrel{\phi_{\ell-1}}\longrightarrow
\cdots \stackrel{\phi_2}\longrightarrow F_1 \stackrel{\phi_1}\longrightarrow F_0
\stackrel{\phi_0}\longrightarrow M\longrightarrow0,
\label{fpr}
\ee
then it is a finite projective resolution and we say it has length $\ell$ (if it does not terminate, then
its length is $\infty$). We note that, without loss of generality, the projective modules $P_0$, \ldots,
$P_{\ell-1}$ in a projective resolution can be replaced by free modules, as shown,
because of the definition of a projective module (the final projective module $P_\ell$ in the sequence
shown may then not be the same one as in the original projective resolution). The length of a free 
resolution (i.e.\ as in eq.\ (\ref{fpr}), but where $P_\ell$ is free) 
is defined the same way. Finally, when $R$ is Noetherian (as our rings are)
and $M$ is finitely generated, each projective or free module in the sequence can be taken to be
finitely generated also.

For any module $M$, there is a minimum
length for a projective resolution, and that minimum is called the right {\em projective dimension} of
the module \cite{passman}. The projective dimension measures how close the module is to being projective;
for example, the projective dimension of a projective (or of a free) module is zero.
Finally, the supremum of the projective dimensions of the modules is called the right {\em global
dimension} of the ring $R$. If all f.g.\ modules have finite right projective dimension,
then we say the ring is right {\em regular} (note that a regular ring could have infinite global
dimension). For any Noetherian ring, the right and left global dimensions are equal. We will sometimes
use the term {\em length} of a module for the minimum length of a {\em free} resolution of the module.

A precursor to the syzygy theorem is the statement that, if a ring $R$ has (right) global dimension
$\cal N$, then the polynomial ring $R[X]$ in one variable has global dimension ${\cal N}+1$
\cite{passman}. As the
global dimension of any division algebra is zero, it follows that the global dimension of the polynomial
rings ${\bf D}[X_1,\ldots,X_d]$ is $d$, where ${\bf D}={\bf R}$, ${\bf C}$, or $\bf H$. Thus this limits
the lengths of minimum projective resolutions, but more is true: it can be proved that any f.g.\
projective module
$P$ over one of these polynomial rings is {\em stably free}, that is there exists a {\em free} module $F'$
such that $P\oplus F'=F$ is free. (Clearly, any stably-free module is projective.) This means that, for
any module over one of these polynomial rings, there is a free resolution whose length is greater
by at most 1 than the projective dimension; that is, a free resolution of length at most $d+1$
\cite{passman}. These statements---that is, that the global dimension is $d$ and that all f.g.\
projective modules are stably free---also hold for the three rings of Laurent polynomials $R_i$ ($i=1$,
$2$, $3$), for any $d$ \cite{mccrob}. This ``weak'' version of the syzygy theorem is sufficient for the
proof of our no-go theorem. Hilbert's syzygy theorem in its original or ``strong'' form says even
more: it says that for $\bf D$ a field (say $\bf R$ or $\bf C$), there is a free resolution of length
$\leq d$ of any f.g.\ module over the polynomial ring; however, we will only rarely require this refinement.

Next we relate these results to algebraic $K$-theory. First, one way to define the 
Grothendieck group $K_0(R)$
for a ring $R$ is as follows \cite{rosenberg}. We begin with the f.g.\ projective
modules over $R$, and take isomorphism classes. A direct sum of f.g.\ projective modules is f.g.\
projective, and is well defined for isomorphism classes. Thus the equivalence classes of the f.g.\
projective modules form an Abelian semigroup, that is a set with an associative, Abelian,
binary operation (which we write as addition). [Indeed they form a monoid, because the
zero module is projective and is the identity element for direct sum.] Given any Abelian semigroup $S$,
there is a universal way to turn it into an Abelian group $G$, called group completion or the Grothendieck
construction. $G$ can be defined as the Abelian group that has one generator for each element of the
semigroup, and relations that state that if $x+y=z$ in $S$ (for elements $x$, $y$, $z\in S$), then
the corresponding generators in $G$ obey the same relations. In particular, if there is an identity
element in $S$, its image
in $G$ is the identity. [$G$ can also be defined as a group of pairs of elements of $S$, obeying some
relations such that a pair $(x,y)$ gives a meaning to the difference $x-y$ which was not in general
defined in $S$,
similar to the usual constructions of the integers from the natural numbers (with addition as the
operation), or of the non-zero rational numbers from the non-zero integers (with multiplication as the
operation).] Applying this definition to the semigroup of equivalence classes of f.g.\ projective
modules over the ring $R$ yields $K_0(R)$.
The equivalence relation involved in passing to the $K_0$ group can be identified as {\em stable
isomorphism} \cite{milnorK,weibelK}: two f.g.\ projective modules $P_1$, $P_2$ map to the same element
in $K_0(R)$ if and only if there
is a free module $F$ such that $P_1\oplus F\cong P_2\oplus F$.
We may note here that if the definition is applied to the case of the ring $C_{\bf C}(B)$ on a
compact Hausdorff space $B$, the result
is isomorphic to the usual topological $K$-theory group $K^0(B)$,
while when $B$ is also Real, for the rings $C_R(B)$, $C_Q(B)$ it produces the
corresponding groups that we denote $KR^0(B)$ (following Atiyah \cite{atiyah}) and $KQ^0(B)$ ($\cong
KR^{-4}(B)$ in Atiyah's notation), respectively; these classify stable isomorphism classes of
respective types of finite-rank vector bundles over $B$, relevant to classes A, AI, AII on putting $B=T^d$.

The group $K_0(R)$ for the polynomial rings $R_i$ is not sufficient for our purposes, because the
module generated by a set of compactly-supported Wannier-type functions is not in general projective.
The generators of the module are supposed to have the property
that, when evaluated as vectors for any $\bk$ (that is, for any set of $X_\mu$ such that $|X_\mu|=1$ for
all $\mu=1$, \ldots, $d$), they span a subspace of ${\bf C}^n$ of rank $m$ [these become ${\bf C}^{2n}$
and rank $2m$ (over $\bf C$) in the case of class AII]. This property of the vector-valued functions on
the torus $|X_\mu|=1$ does not imply much about their behavior at other $X_\mu$, and though the module is
finitely generated, it seems unlikely to be projective in general. For example, consider a submodule of the
rank-one free module $R_i$, which for us is the case $m=n=1$; such a submodule is a (right) ideal in 
$R_i$. Suppose further that $R_i$ is commutative ($i=1$ or $2$). In one variable ($d=1$), the polynomial 
ring $R={\bf F}[X]$ (${\bf F}={\bf R}$ or $\bf C$) is a principal ideal domain (PID), that is, all ideals 
are generated by a single element, so the module is free of rank one. But for $d>1$, such polynomial 
rings $R$ are not PIDs, which implies that there are ideals (submodules of $R$) that cannot be generated by 
a single element. If the polynomials $f$, $g$ are two generators, then there is a linear relation with 
coefficients in $R$ that they obey (namely, $fg-gf=0$), and so the module is not free. But for these 
rings all f.g.\ projective modules are free (Serre's problem, solved independently by Quillen and Suslin 
\cite{passman}), and hence these f.g.\ modules cannot be projective. We can obtain examples for larger 
values of $m$ and $n$ by taking the direct sum of one of these modules with, for example, a free module.

For this reason, we must work with a larger category of modules. We consider the
category of all f.g.\ modules over $R$, together with all the homomorphisms between them. In this 
setting, there is a further Grothendieck group $G_0(R)$, defined as follows 
\cite{passman,mccrob,rosenberg}. (Its use is not essential to the proof, however it can be viewed 
as providing an ``upper bound'' on the classification of the modules of interest.) It is constructed 
from generators, one corresponding to each isomorphism class of f.g.\ modules (call the class $[A]$ if
it contains the module $A$), and relations $[A]+[C]=[B]$ if there is a short exact sequence as in
(\ref{short_exact}) connecting $A$, $B$ and $C$ (again, this is well-defined for isomorphism classes).
Other categories of modules can be handled in the same way, provided they ``possess exact sequences''
\cite{rosenberg,weibelK}; in particular, the category of f.g.\ projective modules can also be treated in
this way, and the result is the same group $K_0(R)$, essentially because short exact sequences of
projective modules split, so $B\cong A\oplus C$. As the f.g.\ projective modules form a full subcategory
of the category of f.g.\ modules, $K_0(R)$ is a subgroup of $G_0(R)$. We note that in $G_0(R)$ or $K_0(R)$
we can also write under the same conditions $[C]=[B]-[A]$, an ``alternating sum'', and that similar
forms apply for the class $[M]$ of the module $M$ in a resolution like (\ref{fpr}) of any finite length,
by iteration of this formula for a short exact sequence. As an example, for the sequence (\ref{fpr}),
we have $[M]=\sum_{i=0}^{\ell-1}(-1)^i[F_i] + (-1)^{\ell}[P_\ell]$.

The statements above about the syzygy theorem now translate into statements about these groups. First,
for any right regular ring $R$, every f.g.\ module $M$ possesses a projective resolution of
finite length. This in effect reduces questions about the structure of $M$ to questions about the
structure of the f.g.\ projective modules (including free modules) in the resolution. In particular, it was
proved by Grothendieck in this manner that $G_0(R)\cong K_0(R)$ \cite{weibelK}; note that what happened
here was that the class (group element) $[M]$ in $G_0(R)$ can be computed as a finite alternating sum of
the classes $[P_i]$ of the f.g.\ projective modules in the finite-length resolution, and these classes
all lie in the subgroup $K_0(R)\subseteq G_0(R)$, and hence so does $[M]$. (This form of calculation based
on the syzygy theorem will recur in the arguments for every symmetry class.) Second, if every f.g.\
projective
module over $R$ is stably free---that is, stably isomorphic to a free module---then it follows immediately
that $K_0(R)$ is generated by the free module $R$, and so $K_0(R)= {\bf Z}$. (Actually, this
also requires that $R$ has the ``invariant basis property'' that for f.g.\ free modules,
$R^n\cong R^{n'}$ implies $n=n'$, which holds for nonzero right Noetherian rings \cite{mccrob,weibelK}.
Then the free modules indeed generate a copy of $\bf Z$ in $K_0(R)$.) Note
that we generally view $K$ and $G$ groups additively, and so also a direct product of groups will be written
as a direct sum, for example ${\bf Z}\oplus {\bf Z}$, because it is additive, and because such a group
is in a natural way a module over the integers, so it is a genuine direct sum. Such a form will also be 
written $2.{\bf Z}$, and similarly for $k.{\bf Z}$ for a positive integer $k$.

These results then imply that for the polynomial rings $R_i$, $G_0(R_i)= {\bf Z}$. This gives the
classification of {\em all} f.g.\ modules over $R_i$ up to the equivalence used in the Grothendieck
construction. (When dealing with more general categories than the category of f.g.\ projective modules,
this relation is no longer stable equivalence.) Within this classification, this result says that
the modules are effectively trivial, as they are described by a single invariant, which corresponds to
the rank of a free module (the invariant is the alternating sum of the ranks in a free resolution of
the module). We would like to relate this to the class of the vector bundle that the elements of one 
of our modules span, within the topological classification of vector bundles of the appropriate type. 
To this end, we can map a module into a vector bundle over $T^d$, and map the corresponding Grothendieck 
$K_0$ groups, in two
steps. For the first step, the rings of polynomials ${\bf D}[X_\mu]$, ${\bf D}={\bf C}$, ${\bf R}$, or
$\bf H$, (and also the corresponding rings of Laurent polynomials ${\bf D}[X_\mu^{\pm 1}]$), can be
embedded into the rings $C_i$ of continuous functions on $T^d$ by evaluating the indeterminates $X_\mu$
as complex numbers with $|X_\mu|=1$, as already discussed. Given a module over one of the polynomial
rings, this produces a module over the corresponding ring of functions, and is an instance of the
change-of-rings functor that can be given formally by tensor product with the latter ring (see Section
\ref{background}). The functor maps f.g.\ projective modules to f.g.\ projective modules, so it gives
a well-defined homomorphism of
groups from $K_0(R_i^{(d)})$ to $K_0(C_i^{(d)})$. Free modules clearly map to trivial
vector bundles, and our $K_0(R_i)={\bf Z}$ maps to the $\bf Z$ in the $K$ group of the vector bundles
that describes the rank and is exemplified by the trivial vector bundles; that is, the homomorphism of
$K_0$ groups just mentioned is injective (one to one). This may suggest that the
general f.g.\ modules over $R$ map in some sense to trivial vector bundles.

The last statement, however, is too naive. For a module over the polynomial ring $R_i$ that is not
projective, its image under the change of rings may not even be a vector bundle (because it is not
projective as a module over the ring of continuous functions). The way this can happen is that the
elements in the module do not span a vector space of full rank $m$ at some points of the Brillouin
torus $T^d$; then vector-valued functions in the module over the ring of continuous functions have the
same property at that point, and we do not have a projective module or a vector bundle. Hence there is
no natural functor from the category of all f.g.\ $R_i$ modules to that of f.g.\ projective $C_i$ modules,
and neither do we wish to begin discussing the category of all f.g.\ $C_i$ modules and $G_0(C_i)$
(and likewise for higher $G$ and $K$ groups relevant to later sections).
Thus it is {\em crucial} that we want to classify, not all f.g.\ modules (over $R_i$), but only
the modules that have the completeness property of Wannier-type functions; the latter property, 
by definition, gives us a vector bundle (projective module) as the image under the change-of-rings functor. 

The basic idea with which to complete the proof, as in DR, is to use the finite-length free resolution of 
the $R_i$-module, and map it onto a similar sequence of $C_i$-modules. Each free $R_i$-module in the 
resolution maps to a free $C_i$-module of the same rank, but there is the question of showing that the 
resulting sequence
(which has the same length) is actually exact. If it is exact, then it gives a free resolution of the 
$C_i$-module, which is the (space of sections of the) bundle of interest. The $K_0$
class of the latter is then given by the alternating sum of those of the free modules in the resolution, 
showing that the $K_0$ class of the bundle is that of a trivial bundle. Hence essentially the only 
remaining point to prove is that the sequence of $C_i$-modules is exact. We note that this requires 
proof because exactness of a sequence of vector bundles means exactness of the maps of the fibres at 
each $\bk$. The maps $\phi_i$ are described by the same $n_i\times n_{i-1}$ matrices 
with entries in $R_i$ as in the free resolution of $R_i$-modules (we put $n_{-1}=n$), and so the composites 
$\phi_i\circ\phi_{i+1}=0$. The issue is whether ${\rm im}\, \phi_{i+1}$ is {\em onto} the $\ker \phi_i$; 
this could fail to hold when evaluated at some $\bk$, even though the sequence over the
polynomial ring is exact. In DR this was proved using the notions of analytic polynomial
bundles. In the present setting of polynomially-generated bundles, a more direct argument using less 
structure seems appropriate, and is given here to make the argument self-contained and perhaps simpler.

The argument proceeds, as in DR, by starting at the right and working back up the sequence. To begin, 
our module $M$ (which we view as a submodule of a free module $C_i^n$), is polynomially generated and 
so projective as a $C_i$-module, and also is the image of the map $\phi_0$ from the free module 
$F_0=C_i^{n_0}$ onto $M$. The kernel of $\phi_0$ is a complex vector space of dimension $n_0-m$ 
[$2(n_0-m)$ for $i=3$] at all (real) $\bk$, by the rank-nullity theorem of linear algebra, and so forms 
a vector bundle. We must show that ${\rm im}\, \phi_1$ spans $\ker \phi_0$ at all (real) 
$\bk$. Because $\phi_1$ is a matrix of polynomials, this means showing that $\ker \phi_0$ is 
itself polynomially generated (the generators are the columns of $\phi_1$.) 

Now studying the kernel of $\phi_0$ means solving a system of $n$ homogeneous linear equations in $n_0$ 
unknowns. For equations with coefficients in a division ring, this can be done
using Gaussian elimination, even in the non-commutative case \cite{artin}. (The approach used in DR 
can be viewed similarly also.) The result of the algorithm
is expressions for $n_0-m$ linearly-independent (over the division ring) vectors in the kernel of the 
linear map. These expressions are the result of
a finite number of arithmetical operations in the division ring, including division by a number of 
``pivots'' \cite{strang}; it is of course important that the latter are invertible (i.e.\ non-zero) 
in the division ring. Our rings $R_i$ are not division rings, but each can be ``completed'' to a 
division ring of ``fractions'' or ``quotients'' by including an inverse for every non-zero element. 
The resulting rings, say $D_i$, with $R_i\subseteq D_i$, consist of all finite linear combinations 
of elements of $\bf R$, $\bf C$, or $\bf H$ for $i=1$, $2$, $3$, respectively, with coefficients that 
are now {\em ratios} of polynomials in $X_\mu$ with real coefficients, in which the denominator must 
not be the zero polynomial. For the commutative cases $i=1$, $2$, $D_i$ is the familiar field of 
rational functions, but the non-commutative $D_3$ is likely less familiar (see Ref.\ \cite{goodwar} 
for general discussion). Note that the inverse in $D_3$ of an element
$r$ of $R_3$ can be expressed in a similar way as for quaternions, as an element of $R_3$
divided by a real polynomial, that is, by an element $|r|^2$ of $R_2$. If $r$ is expressed as a
$2\times 2$ matrix, $|r|^2$ is the determinant, and is a sum of four squares of polynomials, each with
real coefficients; crucially, it vanishes as a polynomial in $R_2$ only when $r=0$ in $R_3$. (We remark 
that for classes A, AI, and AII, the matrix $g_\bk$ used in the definition of a TNS, as in Sec.\ 
\ref{sec:nogo}, has entries in $D_i$.) The Gaussian elimination algorithm can be
carried out in $D_i$, and the resulting solutions form a set of $n_0-m$ $n_0$-component vectors with
entries in $D_i$, and are linearly independent over $D_i$. Finally, we can multiply each $n_0$-component 
vector by a common denominator of its entries \cite{goodwar} to obtain
vectors with entries in $R_i$. These must be linearly independent over $R_i$, because if not then the
original vectors in $D_i^{n_0}$ would be linearly dependent over both $R_i$ and $D_i$.

In slightly more detail, Gaussian elimination in $D_i$ recursively reduces the $n\times n_0$ matrix 
of $\phi_0$, which initially has entries in the subring $R_i$, to echelon form \cite{strang}: 
all rows below the $m$th are zero, and (after permuting the columns, i.e.\ the unknowns, if necessary) 
the top left $m\times m$ block is upper triangular with the pivots, which are non-zero elements of $D_i$, 
on the diagonal. A linearly-independent set of solutions in $D_i^{n_0}$ to the homogeneous equations 
is obtained using the $n_0-m$ standard basis vectors (with a single $1$, and other entries zero) for the
$n_0-m$-dimensional subspace in which the first $m$ entries of the vectors are zero. The denominators 
of the entries of these vectors are products of elements of $R_i$ that are also numerators of pivots.

Now we must investigate the finiteness and linear independence over $\bf C$ of these solutions 
when evaluated in the neighborhood of some $\bk_0$. By hypothesis, we can assume that we have obtained 
an echelon matrix in which, when evaluated at $\bk_0$, the pivots are invertible, meaning non-zero 
for $i=1$, $2$, and invertible as $2\times 2$ matrices for $i=3$. (The denominator of a pivot cannot 
vanish at $\bk_0$, because in Gaussian elimination starting with a matrix with entries in $R_i$, the pivots
are produced in a sequence $1$, $2$, \ldots, $m$, and the denominator of an entry in the $j$th row of 
the echelon matrix can only vanish at $\bk_0$ if one of the pivots from a stage earlier than $j$ vanishes 
at $\bk_0$.) This ensures that the rank of $\phi_0$ evaluated at $\bk_0$ really is $m$, as assumed. 
Then our set of $n_0-m$ vectors in $D_i^{n_0}$ that
span $\ker \phi_0$ are finite when evaluated at $\bk_0$, and still linearly independent.
By continuity of polynomial functions, this is also true in some neighborhood of $\bk_0$. (If
expressed as vectors in $R_i^{n_0}$, they may not be linearly independent when evaluated at some
$\bk$ outside a neighborhood of $\bk_0$.) Thus the dimension of the space of solutions is locally
constant and equal to $n_0-m$. When the common denominator of each vector is removed, they lie in 
$\ker \phi_0$ viewed as a map of $R_i$-modules, and the sequence of modules over $R_i$ was exact,
so these vectors also lie in ${\rm im}\, \phi_1$. Hence ${\rm im}\, \phi_1$ spans 
$\ker \phi_0$ when evaluated at any (real) $\bk$. This means the sequence of $C_i$-modules is exact at 
$F_0$, or in other words that the corresponding sequence of vector bundles 
is exact at $F_0$. The argument can now be iterated to show that the whole sequence of vector bundles 
is exact; it can also be called a free resolution.

Exactness of the sequence of vector bundles, or in other words the free resolution of the projective $C_i$ 
module, now allows us to apply the (at this stage, algebraic) $K_0$ functor to the modules in the sequence,
as they are all projective. As explained above, the $K_0(C_i)$ class of our module $M$ (or corresponding 
bundle) is the alternating sum of those of the free $C_i$-modules in the sequence (the same alternating 
sum as for the free resolution over $R_i$). The classes of
the free $C_i$-modules lie in the image of the injective homomorphism from $K_0(R_i)$ to $K_0(C_i)$ that 
is induced from the change of rings map. Alternating sums of these classes also lie in the same group. 
Hence the $K_0$-class of the vector bundle that we have obtained is that of a free $C_i$-module, and 
$K_0(R_i)$ classifies the polynomially-generated bundles. In each of the three cases, this $K_0(R_i)$ 
group is the group of integers $\bf Z$, and the integer-valued invariant associated with the bundle can be 
identified as its rank (it will always be positive). (The analysis for the remaining symmetry classes 
will also follow a similar path, which is similar to the method for proving that $G_0(R_i)\cong K_0(R_i)$.)

In DR the conclusion of the no-go theorem (or its extension to compactly-supported Wannier functions) 
was stated as saying that the polynomially-generated bundle (or in particular, the analytic polynomial 
bundle) obtained must be trivial (as a complex vector bundle, as only class A was considered). In fact, 
in general the proof given there or here only establishes that the bundle is {\em stably} trivial. We 
define stably trivial to correspond to the definition of a stably-free module, given above, on passing to 
the space of sections: a bundle $E$ is stably trivial if there are trivial bundles $F$, $F'$ such that 
$E\oplus F\cong F'$. Indeed, stable triviality of our bundle follows directly from the existence of 
a finite-length free resolution: as the image of each map is a projective module, the sequence splits as 
a direct sum at each term. Then $M\oplus\ker \phi_0\cong F_0$ is free; $\ker \phi_0$ might not be free, 
but we can apply $\oplus\ker \phi_1$ to both sides, to obtain the free module $F_1$, and so on.
This process terminates after a finite number of steps, and the result $M\oplus F_1\oplus\ldots \cong 
F_0\oplus F_2\oplus\ldots$ [ending with $P_\ell$ (free) and $F_{\ell-1}$ on the two sides] shows the stable 
freeness of the module (i.e.\ stable triviality of the bundle). (On mapping to $K_0$ classes, this 
produces an expression equivalent to the one for $[M]$ as an alternating sum.)

It was assumed in the main proofs in 
DR that stably trivial is the same as trivial, when dealing with the (split) exact sequence of bundles. 
But in general it does not seem that stably-trivial bundles are always trivial; there are counterexamples at
least for real (not Real) vector bundles \cite{huse}. However, for $d\leq 3$, complex vector bundles over
any $d$-dimensional manifold $B$ can be reduced (i.e.\ are isomorphic to) to a direct sum of a rank-one (or
``line'') bundle and a trivial bundle \cite{huse}, and the complex line bundles in any dimension $d$
are classified up to isomorphism by their first Chern numbers (see e.g.\ Ref.\ \cite{weibelK}, page 45).
(This notion of ordinary isomorphism of complex vector bundles describes our problem, i.e.\ rank-$m$
complex vector sub-bundles of a trivial rank-$n$ bundle, in the case of $m$ fixed and $n$ sufficiently
large. For general values of $m$ and $n$, a finer classification is possible; see e.g.\ Ref.\ \cite{kz}.)
Hence for complex vector bundles in $d\leq 3$, stably trivial and trivial are the same. DR also established
triviality directly in some special cases. But in general, the conclusion of our analysis should be stated
as the stable triviality of the bundles for classes A, AI, and AII. This stable triviality is what
$K$-theory deals in, and for class A with $B=T^d$ corresponds also to the vanishing of all Chern classes
of a bundle.

To avoid a possible confusion, we should mention that when the polynomial sections fail to span the
space of rank $m$ at some point in the Brillouin torus, it may still be possible to produce a non-trivial
vector bundle, even though the Grothendieck group of all f.g.\ modules $G_0(R)={\bf Z}$. The
idea is to allow sections obtained from those mentioned already by using functions that tend to infinity
at the points in question; the pseudo-sections obtained that way may span a vector bundle (i.e.\ be
continuous as vector-valued functions of $\bk$, that span a rank-$m$ subspace). This is the phenomenon
discovered in DR \cite{dubail} and Ref.\ \cite{cirac}, which can lead to a non-trivial vector bundle in
this manner. However,
such a vector bundle is necessarily non-analytic, and the reason for the non-triviality of the bundle
(contrary to the classification above) is the use of non-continuous (diverging)
coefficients when the vector bundle was obtained from the polynomial sections.

This essentially concludes the argument for these symmetry classes, but we have not yet mentioned the
second step in relating the algebraic classification of modules over polynomial rings to the topological
classification of vector bundles over the Brillouin torus. This step is the passage from the algebraic
to the topological classification in the case of projective modules over a ring of continuous functions
on $T^d$, which correspond to vector bundles. While this requires more explanation in some cases (see
the following section), in the present case the result of the algebraic classification [e.g.\
$K_0(C_{\bf C}(B))$ in the complex case] is already a discrete group, which is just isomorphic to
the topological $K$-theory group [$K^0(B)$ in the complex case] \cite{milnorK}, so there is nothing more
to do.

We mention here some consequences of the strong form of the syzygy theorem for low-dimensional versions
of our problems, in terms of the nature of the modules over the appropriate polynomial ring. We begin 
with classes A and AI. First, a set of compactly-supported Wannier-type
functions generate a module which we will now call $M'$ over one of the commutative polynomial rings 
$R_i$ ($i=1$, $2$), which is a submodule of a free module $F_0$ of rank $n$ over $R_i$. The quotient 
module of $F_0$ by $M'$ is a module $M$, and so in
the resolution (\ref{fpr}) $M'$ is the image
of $\phi_1$. In one dimension ($d=1$), the syzygy theorem says there is a length 1 free resolution of
$M$, and so $M'$ is actually a free module. (To be precise, we have used here not only the {\em existence}
of a length $d$ free resolution, but the fact that any {\em given} projective resolution can be truncated
to one of length at most $d$, because it splits \cite{passman,mccrob}, together with the fact that any
projective module is free. The result for $d=1$ also follows
directly from the fact that the polynomial ring in one variable is a
PID \cite{passman,rosenberg}.) For the polynomial bundles, relevant to TNSs, the module of interest
(say $M''\subseteq F_1$) is defined as the solutions to polynomial equations, and so is the kernel
of a map $\phi_1$, again into a free module, and the cokernel of $\phi_1$ can be taken as the module
$M$. In one dimension, this sequence is longer than required by the syzygy theorem, and so again splits:
$M''$ is a direct summand in $F_1$, and so is projective, and actually free in the real and complex cases.
Put another way, the preceding result applies {\it a fortiori} to $M''$. For $d=2$, we see from the syzygy
theorem that $M''$ is a free module; this extends the result of DR that for a rank 1 vector bundle, the
module is free in any dimension. In general, the module $M'$ in the case of compactly-supported Wannier-type
functions has length $d-1$, while in the case of a TNS the module $M''$ corresponding to the filled-band
bundle has length $d-2$ for $d\geq 2$ (here we have corrected some misstatements in DR about the minimum
length of the free resolutions that did not invalidate any results). For class AII, the polynomial ring
${\bf H}[X_\mu]$ is not commutative. The same statements as before hold in one dimension, however
stably-free projective modules over ${\bf H}[X_1,X_2]$ are not always free
\cite{passman}, so the ``strong form'' of the syzygy theorem does not hold for $d\geq 2$. Hence, when
$d=2$, $M''$ does not have to be free, but must be stably free and hence projective.

\section{Chiral symmetry classes AIII, BDI, CII: $K_1(R)$}
\label{chiral}

In this section, we turn to the classes with so-called chiral symmetry. These are the chiral unitary
ensemble, or class AIII, chiral orthogonal class BDI, and chiral symplectic class CII. For these cases,
we will handle the three classes mostly in parallel.

\subsection{Chiral symmetry classes}

A typical way for chiral symmetry
to arise in a tight-binding model is when there are two sublattices, say $A$ and $B$, and the only possible
hops are from one sublattice to the other. Then in a basis in which the indices for sites and orbitals
are partitioned into the two subsets corresponding to the two sublattices, the Hamiltonian has the block
off-diagonal form
\be
{\cal H}=\left( \begin{array}{cc}0&h\\h^\dagger&0\end{array}\right),
\ee
with square blocks (in cases where the off-diagonal blocks are not square, there are zero-energy
states in the spectrum; we do not consider this as we wish to discuss only topological phases), and $h$ is
an arbitrary matrix with complex entries. (This most general case defines class AIII; we discuss the
other chiral classes afterwards.) In the translation-invariant case on a lattice, with $2n$ orbitals per
site ($n$ assigned to each ``sublattice''; note that in our hypercubic lattice models, they actually all
sit on the same lattice sites), the Hamiltonian in $\bk$ space has the similar form
\be
{\cal H}_\bk=\left( \begin{array}{cc}0&h_\bk\\h_\bk^\dagger&0\end{array}\right),
\ee
where the blocks are now $n\times n$. To ensure a gap in the energy spectrum, we assume that $h_\bk$
is non-degenerate at all $\bk$. The chiral symmetry acts as multiplication by $1$ on all orbitals
on the $A$ sublattice, and by $-1$ on the $B$ sublattice, that is by conjugating the Hamiltonian by
$\Sigma_z$: $\Sigma_z{\cal H}_\bk\Sigma_z=-{\cal H}_\bk$, which forces it to have the above form.

The energy eigenvalues of ${\cal H}_\bk$ come in plus-minus pairs, determined by the square roots of the
eigenvalues of ${\cal H}_\bk^2$, and so by the singular values of $h_\bk$ (the positive square roots of
the eigenvalues of $h_\bk^\dagger h_\bk$). A complete orthonormal set of eigenvectors can be written as
the columns of a matrix of the form
\be
\frac{1}{\sqrt{2}}\left( \begin{array}{cc}I_n&I_n\\U&-U\end{array}\right),
\label{2sublat}
\ee
where $I_n$ is the $n \times n$ identity, $U=U_\bk$ is a unitary matrix function of $\bk$, and the first
$n$ columns are basis vectors for negative-energy (filled) bands, and the others for positive-energy
(empty) bands. The chiral symmetry acts by multiplication by $\Sigma_z$ from the left, and exchanges
the eigenvectors corresponding to $E_\bk$ and $-E_\bk$.

For the chiral orthogonal (chiral symplectic) version, we also impose time-reversal symmetry as
discussed in Section \ref{aiaii}, with $\widehat{T}^2=+1$ ($-1$), which in $\bk$-space implies
that the entries of ${\cal H}_\bk$ are Real (Quaternionic), and likewise $U_\bk$ must be unitary (at
each $\bk$) with Real (Quaternionic) entries. Consequently, the vector bundles of rank $2n$ with chiral
symmetry are classified topologically by the homotopy classes of maps (without basepoints) of $T^d$
into $U(n)$; in the Real (Quaternionic) case, the maps involved also respect the involution
$\bk\to-\bk$ which acts as $\widehat{T}$ on the entries of the unitary matrix $U$ in either case. (For
the Quaternionic case, the ranks over $\bf C$
mentioned are doubled due to spin; we will usually not mention this, just as if we describe the rank
over $\bf H$ for a quaternionic vector space.) We discuss the precise (basis-free) meaning of this
statement in the next section. The limits as $n\to\infty$ of these groups of homotopy classes of maps
give the topological $K$-groups $K^{-1}(T^d)$, $KR^{-1}(T^d)$, and
$KQ^{-1}(T^d)\cong KR^{-5}(T^d)$ \cite{atiyah,atiyahbook,karoubi} which classify the topological classes
of band structures in these symmetry classes. This definition for
$K^{-1}(B)$ is equivalent to another definition as $\widetilde{K}^0(S(B^+))$, where $\widetilde{K}^0$ is a
``reduced'' $K$-group, $S$ is ``reduced suspension'', and $B^+$ is $B$ with a disjoint basepoint
adjoined to it \cite{atiyahbook,karoubi}; the equivalence arises because the map into unitary complex
matrices determines the ``clutching function'' used to construct a rank-$n$ bundle over $SB$
\cite{atiyahbook,karoubi}.

For a set of compactly-supported Wannier-type functions to respect the chiral symmetry, we only require
that they come in pairs related by the symmetry transformation (multiplication by $\Sigma_z$). Since we
allow overcomplete sets, this means that even if we begin with a set that does not respect the symmetry,
we can simply include all polynomial vectors obtained by the symmetry action as generators.

\subsection{Classification by $K_1$}

First, we will unpack the implications of our assumptions about the compactly-supported Wannier-type
functions in these classes. At each $\bk$ there is a subset of these functions (polynomial $2n$-component
vectors in $\bk$ space) that span the fibre of the rank-$n$ filled-band bundle, and their counterparts 
(obtained by applying $\Sigma_z$) span the empty-band bundle. Together, they span at each $\bk$ the trivial
rank-$2n$ vector bundle of the total system. Clearly, by restricting to the first $n$ components, we
obtain generators for a module (always over one of the polynomial rings $R_i$ in this paragraph) in
the sublattice $A$
orbitals, and restricting to the last $n$ components gives a
set generating a module in the sublattice $B$ orbitals. (Instead of restricting, we could take the sum
and difference of the pair of corresponding functions.) The latter sets of orbitals give in $\bk$-space
two trivial vector bundles of rank $n$ over $T^d$, which are two orthogonal subbundles of the total rank
$2n$ trivial vector bundle, and the
overcompleteness of the generating sets of sections implies that their restrictions to the two subbundles
span each of them at all $\bk$. At the same time, there is a correspondence
between the fibres of the two (sublattice) subbundles, which is an invertible linear map between the
fibres, exactly like that defined in (\ref{2sublat}), except that here we allow $U$ to be a general
invertible matrix (with entries that are complex, Real, or Quaternionic functions, depending on the
symmetry class). Given any element of the module (i.e.\ a polynomial section)
associated to sublattice $A$, this map gives an associated element of the module associated to sublattice
$B$, and there is an inverse map. Hence we obtain an invertible homomorphism between the two $R_i$-modules.

We point out that for two free modules of the same rank over a ring $R$, presented as $R^n$,
an invertible map from the first to the second is equivalent to an invertible matrix with entries in
$R$, because each generator of the first, represented by a column vector $(0,\ldots,1,0,\ldots,0)^T$ with
a single nonzero entry, must be mapped to a vector with entries in $R$. For modules that are isomorphic to
a free module, an invertible map (or isomorphism) can be represented this way, up to changes of basis on
the two free modules. But maps between general modules need not have this form. For modules that are
submodules of a free module, like many of those we study, a map from one to another can be expressed as
a matrix because of the embeddings of the modules in the free modules, but the entries of the matrix need
not be elements of
$R$; it is only necessary that it map an element of the first module to an element of the second, not
that it map the standard basis vectors to vectors with entries in $R$. That is, it does not have to
be a homomorphism of the first free module into the second at all, and so even though invertible on the
submodules, it does not have to be an invertible map of the first free module to the second. On the
other hand, if an element of the
first module is now expressed as a linear combination of generators (with coefficients in $R$), and so is
its image in the second, then the map determines a matrix with entries in $R$ (because the generators
express each module as the image of a free module, represented by matrices with entries in $R$). However,
the expression as a linear combination of generators in either place need not be unique, and consequently
the matrix of the map, or of its inverse, need not be unique.

This shows that the basic problem we need to study is posed as two submodules of free modules of rank
$n$, with an invertible map (isomorphism) between the submodules. This will lead us to the algebraic
$K_1$ groups of the various rings. A first basic point is that if, naively, we attempt to classify
isomorphisms $\alpha:M_1\to M_2$ (up to isomorphism) between two given modules $M_1$, $M_2$, the result
will be trivial, because we can simply compose the given isomorphism with an automorphism of $M_2$
(i.e.\ an isomorphism to itself; it could be viewed as a change of basis if $M_2$ is free and expressed
as $R^n$), and so obtain any isomorphism from $M_1$ to $M_2$. To obtain anything non-trivial from the
given data, we will have to compare different isomorphisms between the given modules. Thus if we pick
an isomorphism, say $\alpha_0:M_1\to M_2$, and use it as a reference with which to compare other
isomorphisms $\alpha$, then this is equivalent to studying the {\em automorphism} $\alpha_0^{-1}\alpha:
M_1\to M_1$, which is not necessarily a trivial problem. (Indeed, when we discussed the basic band-structure
problem above, we described it as classifying $U$, the unitary matrix functions of $\bk$. There we
referred to some fixed bases for the two subspaces on the two sublattices. This choice of two bases sets
up a reference isomorphism, so that the present formulation is equivalent to what was used there.) Using
henceforth the notation $\alpha:M\to M$ for an automorphism on $M$, the effect of a further
change-of-basis automorphism (say $\alpha'$) on $M$ is to conjugate $\alpha$ with $\alpha'$: $\alpha\to
\alpha'^{-1}\alpha\alpha'$. This cannot reduce $\alpha$ to the identity $\rm id$ unless $\alpha={\rm id}$
to begin with. It is also clear that if we want to classify automorphisms of a module in a meaningful way,
then those that differ by conjugation should be viewed as equivalent.

We are now ready to introduce the definitions of the Bass-Whitehead group $K_1(R)$ for a ring $R$, 
or in fact $K_1({\cal C})$ for any category $\cal C$ with exact sequences \cite{rosenberg} (going straight 
to the more general form
this time). We consider pairs $(M,\alpha)$ where $M$ is an object in the category (we can think of $M$ as a
module over $R$), and $\alpha$ is an automorphism of $M$. We construct a type of Grothendieck group for
these pairs, as an Abelian group with a generator $[(M,\alpha)]$ for each pair $(M,\alpha)$,
subject to the following two types of relations: (i) for two automorphisms $\alpha$, $\beta$ of $M$,
\be
[(M,\alpha)]+[(M,\beta)]=[(M,\alpha\beta)],
\ee
(so composition of automorphisms gives addition of elements in $K_1$, with in particular $[(M,{\rm id})]$
as the zero element of the group for all $M$; note this relation also shows that $\alpha\beta$ and
$\beta\alpha$ give the same element, since the group operation $+$ is Abelian); and (ii) if there is a
commutative diagram in the category with exact rows
\be
\begin{array}{rrcrcrcrl}
0&\longrightarrow &\hphantom{\alpha_1}M_1&\hphantom{\alpha_1}\longrightarrow&
\hphantom{\alpha_1}M_2&\hphantom{\alpha_1}\longrightarrow&
\hphantom{\alpha_1}M_3&\longrightarrow& 0\\
&&\alpha_1\downarrow&&\alpha_2\downarrow&&\alpha_3\downarrow&\\
0&\longrightarrow &\hphantom{\alpha_1}M_1&\longrightarrow& \hphantom{\alpha_1}M_2&\longrightarrow&
\hphantom{\alpha_1}M_3&\longrightarrow& 0,
\label{comm_diag}
\end{array}
\ee
where $\alpha_1$, $\alpha_2$, $\alpha_3$ are automorphisms of $M_1$, $M_2$, $M_3$, respectively, then
\be
[(M_2,\alpha_2)]=[(M_1,\alpha_1)]+[(M_3,\alpha_3)].
\ee
The latter implies in particular invariance under conjugation of an automorphism (set, say, $M_1=0$, and
$M_2=M_3$). It also implies (taking $M_1=M_3$, $M_2=M_1\oplus M_3$, $\alpha_2=\alpha_1\oplus \alpha_3$)
that addition of two automorphisms of the same module (as direct sum) is equivalent to composition
(mentioned just now). When $\cal C$ is the category of all f.g.\ right $R$ modules, the resulting group
is called $G_1(R)$. When instead $\cal C$ is the category of f.g.\ projective modules, it is called
$K_1(R)$. Apart from the polynomial rings $R_i$, the preceding definition of $K_1(R)$ can be applied
without change to the case of the rings of continuous functions $C_i$, and of vector bundles or
f.g.\ projective modules over such a ring, to obtain the groups $K_1(C_i^{(d)})$.

For the case in which $R$ is a right regular ring, there is a theorem analogous to that for $K_0(R)$
and $G_0(R)$, namely $G_1(R)\cong K_1(R)$ (again due to Grothendieck \cite{rosenberg}). Once again, 
this is a consequence
of the syzygy theorem, together with a result that given a module $M$ with an automorphism $\alpha$,
there exists a projective resolution with compatible automorphisms of each term (of a form similar to
the commuting diagram above) \cite{rosenberg}.

Now we need to calculate $K_1(R_i)$ for the Laurent polynomial rings $R_i$. (In the present case, one
cannot reduce
the problem to the ordinary polynomials, because multiplying generators of the module by positive powers
of $X_\mu$ cancels out in the automorphism. Indeed, the $K_1$ groups of the two types of rings are
different, as we will see.) We can begin with a simpler looking problem: we suppose that $M$ is a free
module $R^n$ for some $n$. Then an automorphism $\alpha$ is represented by an invertible matrix with
entries in $R$, that is an element of the group $GL(n,R)$. As free modules are projective, $\alpha$ will
give rise to an element in $K_1(R)$, and we know that composition of automorphisms gives composition in the
group, and (hence) that conjugation by an automorphism leaves the element invariant. If $R$ is commutative,
then we know that the determinant of the matrix of $\alpha$ has these properties; here the determinant is
a map onto $R^\times$ (because the matrix must be invertible), viewed as a multiplicative group. [In
general, we might think of a determinant operation on invertible $n\times n$ matrices as defining a
homomorphism of $GL(n,R)$ into an Abelian group, whether or not $R$ is commutative.] Thus for the Laurent
polynomial rings $R_1$ and $R_2$, the determinant map gives (calling the resulting group
$\det R$ temporarily)
\bea
\det R_1^{(d)}& =& {\bf C}^\times \oplus d.{\bf Z}\\
\det R_2^{(d)}& =& {\bf R}^\times \oplus d.{\bf Z}.
\eea
Here we reverted to additive notation for direct products of groups; the integers arise from the exponents
$m_\mu$ in the units. Note that we could write the group of units additively as ${\bf Z}/2\oplus {\bf R}$
instead of the multiplicative ${\bf R}^\times$ (the latter is obtained by using the exponential map). For
${\bf C}^\times$, it can be obtained by applying
the exponential map to the additive group $\bf C$, but note that
this map of course has a kernel, the integer multiples of $2\pi i$. These groups are independent of the
size $n$ of the matrices used; the determinant already gives values in these groups for the $n=1$
case.

For matrices whose entries are quaternions, it is possible to define a determinant, the Dieudonn\'{e}
determinant \cite{artin, rosenberg}. There seems to be no simple algebraic expression for it, but the result
is
\be
\det R_3^{(0)} = {\bf R}^\times_{>0},
\ee
the group of positive real numbers under multiplication. For quaternions, the group of units
${\bf H}^\times$ is not Abelian, whereas $K_1$ and $\det$ should be. The properties of a determinant,
applied to $1\times 1$ matrices, imply (at least if it is independent of $n$) that $\det R$ should contain
a quotient of $R^\times_{\rm ab}$. Here for any group $G$ we define its Abelianization to be
$G_{\rm ab}=G/[G,G]$, where $[G,G]$ denotes the group generated by elements of $G$ that are
group-theoretic commutators, that is $[g,h]=ghg^{-1}h^{-1}$ for $g$, $h\in G$. For quaternions,
the group of units ${\bf H}^\times$ is isomorphic to ${\bf R}^\times_{>0}\times SU(2)$ by using the
norm $|q|$ defined above, so ${\bf H}^\times_{\rm ab}\cong {\bf R}^\times_{>0}$, and the map ${\bf
H}^\times\to {\bf R}^\times_{>0}$ can be represented by the norm map $q\to |q|$. For the Laurent
polynomial ring, the units were described above, and the Abelianization is
\be
R_{3\,\rm{ab}}^{(d)\times}={\bf R}^\times_{>0}\oplus d.{\bf Z}.
\ee
One would expect this group to play the role of the determinant group for this ring. We also note that, if
one decides to represent Quaternionic functions by $2\times 2$ matrices of complex functions, then the 
usual (complex) determinant can be taken, and the values for an invertible matrix are of the form 
$|c|^2\prod_\mu
X_\mu^{2m_\mu}$, where $c\neq 0$ is a quaternion and all exponents are even. Thus the result is effectively
the same, and this gives some justification for identifying $\det R_3^{(d)}=R_{3\,\rm{ab}}^{(d)\times}$
also. (Only our reluctance to take a square root stops us from using this as the basis for a
definition of the determinant for matrices with Quaternionic polynomial entries.)

The group $GL(n,R)$ has an obvious embedding into $GL(n+1,R)$ given by
mapping $n\times n$ matrices to the $n\times n$ top left block, with a $1$ at the bottom right place,
and zeroes elsewhere. The sequence of embeddings allows us to take the direct limit as $n\to\infty$
of these groups,
called $GL(R)$. $K_1(R)$ can in fact be defined as $GL(R)_{\rm ab}$, the Abelianization of $GL(R)$; this
can be shown by using the fact that the projective modules are defined as direct summands in a free
module \cite{rosenberg}. This makes $K_1(R)$ a ``stable'' version of a determinant. (Likewise, the group
$K_0(R)$ can also be defined using idempotent matrices over $R$, that is matrices $p$ with entries in $R$
such that $p^2=p$ \cite{rosenberg}, similar to projection operators
that can be used to define a vector bundle as a subbundle of a trivial vector bundle. In both cases,
these definitions
show that the $K$ groups are the same whether right or left modules are used in the other definitions.)
In the case of a division ring
${\bf D}={\bf R}$, $\bf C$, or $\bf H$, it turns out that $K_1({\bf D})$ is precisely the Abelianized
group of units in each case, that is, as in the results above with $d=0$.

These results leave the question of whether the determinant (when defined) or $R^\times_{\rm ab}$
actually is all of $K_1(R_i)$ for the polynomial rings when $d>0$. This question is answered (affirmatively)
in a different way by a further result. If $R$ is a right regular ring, then $K_1$ of the Laurent
polynomial extension ring $R[t,t^{-1}]$ (with indeterminate $t$) is
\be
K_1(R[t,t^{-1}])\cong K_1(R)\oplus K_0(R);
\ee
this is part of the ``fundamental theorem of algebraic $K$-theory'' proved by Bass, Heller, and Swan
\cite{rosenberg,weibelK}. Because a Laurent polynomial ring is itself right regular, the result can
be applied iteratively, and using the results for $K_0$ and the $d=0$ results for $K_1$ already stated, we
obtain
\bea
K_1(R_1^{(d)})&=&{\bf C}^\times\oplus d.{\bf Z},\\
K_1(R_2^{(d)})&=&{\bf R}^\times\oplus d.{\bf Z},\\
K_1(R_3^{(d)})&=&{\bf R}^\times_{>0}\oplus d.{\bf Z},
\eea
in agreement with the Abelianized groups of units discussed above. Note that, as mentioned already,
these also give the corresponding $G_1$ groups, relevant to the classification problem
in which we are interested. We may mention that under the same conditions $K_1(R[t])=K_1(R)$,
so that $K_1$ for ordinary and for Laurent polynomial rings differ for $d>0$, whereas for the $K_0$
groups they are all the same and independent of $d$.

The presence of continuously-varying factors in these groups may seem surprising to readers used to the
topological classifications of free-fermion topological phases by topological $K$-theory. But there is
a simple way to map the algebraic classification here (essentially based on isomorphisms) into a
topological one (essentially based on homotopies). The continuous factors in the groups above represent
distinctions between automorphisms, already seen for a $1\times 1$ matrix or an element of $R^\times$,
which clearly can be continuously deformed to one another. Thus for a classification up to homotopy
equivalence, we can simply remove the continuous factors, which more formally means we take the quotient
by the path-connected component of the identity element of the group \cite{weibelK}, which is a normal
subgroup. We can also describe this operation as passing to the homotopy set of path-connected components
$\pi_0(K_1)$ of the $K_1$s, which we write as $\pi_0K_1$; this homotopy set inherits a group structure.
Then we obtain
\bea
\pi_0K_1(R_1^{(d)})&=&d.{\bf Z},\\
\pi_0K_1(R_2^{(d)})&=&{\bf Z}/2\oplus d.{\bf Z},\\
\pi_0K_1(R_3^{(d)}) &=&d.{\bf Z}.
\eea

We will now confirm (i) that this classification of modules up to equivalence describes what can be
attained with compactly-supported Wannier-type functions, and (ii) relate this to the topological
classification of all band structures in these three symmetry classes. For (i), it is sufficient to
point out that rank one (or $n=1$) examples exist corresponding to the groups $\pi_0K_1$ just obtained. This
is clear, because we already discussed how the results correspond to the (Abelianized) groups of units
$R^\times_{i\,{\rm ab}}$ ($i=1$, $2$, $3$) of the three polynomial rings. For each choice of a unit,
there is a corresponding vector bundle in the chiral-symmetry class in question (set $U$ equal to the
unit; we
have pointed out already that with a given choice of basis for the free module associated to each
sublattice, the automorphism $\alpha$ is represented by the matrix $U$, which here is $1\times 1$),
and it is immediate that the vector bundles are polynomially generated. These constructions
correspond
to Wannier functions that in position space are simply dimers, with one end on a single site in the
single $A$ orbital, and the other end in the $B$ orbital on a site displaced by $(m_1,\ldots,m_d)$,
the set of exponents of the $X_\mu$ in $\bk$ space. Incidentally, these examples also possess a
flat-band parent Hamiltonian with $h_\bk=U_\bk^{-1}$, provided that in the unit $|c|=1$ [see eq.\
(\ref{monom})]. For (ii), it is also immediate that these examples are non-trivial in the
topological classification of vector bundles of these symmetry classes. The exponents $m_\mu$ are ``winding
numbers'' for the behavior of $U$ in $\bk$ space (i.e.\ when evaluated at $|X_\mu|=1$). Part of the
characterization of vector bundles of these classes in general uses the one-dimensional winding number of
the automorphism
$U$, which is essentially the winding of the determinant, similar to the above discussion. (For the
Quaternionic case, the complex determinant is usually used, leading to the appearance of factors of $2$.)
These uniquely label the ``weak topological insulators'' that arise in dimensions larger than 1 by using
the topology of one-dimensional systems, namely a winding in each of the $d$ directions.

To prove that these results classify all polynomially-generated vector bundles in the chiral symmetry
classes, we will again proceed in two steps, as in the case of $K_0$
in the previous Section; namely, we first consider a map (a functor) from the $K_1$ group of
the polynomial ring $R_i$ to the $K_1$ group of the corresponding ring $C_i$
of continuous functions, and then the map from the latter to the topological $K$-theory groups $K^{-1}(B)$.
The argument does not use the full strength of the $G_1(R)$ groups, since we are not interested in all
f.g.\ modules, but these groups do provide an ``upper bound'' on the groups classifying the modules in
which we are interested: the ones generated by (the Fourier transforms of) a set of compactly-supported
Wannier-type functions.
However, our arguments do use the methods that went into the proof that $G_1(R)\cong K_1(R)$ for the
polynomial rings. Namely, as mentioned above, for any f.g\ $R_i$-module $M$ with an automorphism $\alpha$,
we can set up a resolution that is equipped with an automorphism of the free module at each step, that is
a commutative diagram similar to the sequence (\ref{comm_diag}) above, though possibly longer in the
horizontal direction (this involves the ``resolution theorem'' for $K_1$ \cite{rosenberg}). The resolution
essentially reduces the equivalence class $[(M,\alpha)]$ [in $G_1(R_i)$] to an alternating sum
of the classes [in $K_1(R_i)$] of the automorphisms of the free modules, which is why $G_1(R_i)\cong 
K_1(R_i)$. (The automorphisms of the free
modules in the resolution can be described as matrices with entries in the Laurent polynomial ring; 
we know that the class in $K_1(R_i)$ of one of these is determined by the determinant of the matrix.)

If the module $M$ we begin with is generated by a set of compactly-supported Wannier-type functions, then
we know that the exact sequence of free modules in the resolution becomes an exact sequence of trivial
vector bundles (free modules) when viewed as modules over the continuous functions. Moreover, the
automorphism of each free module is given by a matrix with polynomial entries, and is invertible over
$R_i$, so its determinant is an (Abelianized) unit of $R_i$, and in particular is
therefore nonzero at all real $\bk$, as required for an automorphism of a free module over the ring $C_i$.
(Such an exact sequence of free modules and automorphisms can still involve non-trivial automorphisms of
each free module.) This is an application of the change-of-rings functor (see Sec.\ \ref{background}), 
to the change from $R=R_i$, to $S=C_i$, using the natural embedding; let us denote
these generically by $R\to S$. This maps a free module $R^n$ to
$S^n$, a f.g.\ projective $R$-module to a f.g\ projective $S$-module, and a direct sum to a direct sum.
Consequently, it induces a natural map (a homomorphism) $K_1(R)\to K_1(S)$.
It is important that this is well-defined; it means that free modules and automorphisms that are in the
same class in $K_1(R)$ are also in the same class when mapped to $K_1(S)$. In our case, the former contains
$d$ copies of the integers (we disregard the other part for a moment), as does the latter (the winding
numbers mentioned two paragraphs ago).

We will pause the main argument to address $K_1$ of a ring $S$
of continuous functions briefly. For any commutative ring $S$, such as $C_1$ and $C_2$,
the algebraic $K_1(S)$ has a natural decomposition as a direct sum of $S^\times$ coming from the
determinant, plus in general a remainder called $SK_1(S)$ which can be defined as $SL(S)/[GL(S),GL(S)]$
(here $SL(S)$ is the subgroup of $GL(S)$ of matrices with determinant $1$, and note that all commutators
have determinant $1$). For us, $S^\times$ is the multiplicative group of nowhere-vanishing continuous
functions, so is rather large, but ultimately only its image under the homotopy equivalence that mods
out the connected component of the identity will be of interest, and that leaves only the group of
homotopy classes of such nonvanishing functions (under multiplication). (Here we are referring to the
second step of mapping.) For functions with values in ${\bf C}^\times$ (as both examples are, at least
away from points where $\bk \equiv -\bk$), the homotopy classes are obtained by considering only
functions into $U(1)$, and for $B=T^d$ these are determined only by their winding numbers on going around
the torus in one of the $d$ directions. It is clear that the $K_1(R)$s map onto this group $d.{\bf Z}$ in
each case. It should be similar for the non-commutative rings $R_3$ and $C_3$ also; we can carry through
the argument using the ordinary complex determinant of a matrix of Quaternionic functions with the
quaternions expressed as $2\times 2$ matrices, as before, and so define $SK_1(R_3)$ even in this case.
In the particular case of $R_2$, $\pi_0K_1(R_2)$
has an additional summand ${\bf Z}/2$ due to the sign of the determinant, which also occurs in the
corresponding determinant for $\pi_0K_1(C_2)$.

We now return to the free resolution of our $R_i$-module with automorphism, which we converted (by
change-of-rings $R_i\to C_i$) to a similar resolution by free $C_i$-modules (trivial bundles) of the
bundle (with automorphism) of interest. We see that the
automorphism of the bundle is classified [in $K_1(S)$] by an element that is an alternating sum of
elements of $K_1(R)$, which we can view as a subgroup of $K_1(S)$. Consequently, the possibilities
are classified algebraically by $K_1(R_i^{(d)})$; we know that all of these can be attained in some
rank-one example. This concludes the more constructive part of the argument.

Finally, for the second step, in which one removes the continuous part of the space $K_1(S)$ by passing
to the homotopy group of connected components, Milnor \cite{milnorK} (chapter 7) shows, for any
commutative ring of continuous functions, that the only effect is to remove the part we just discussed
as the continuous part of the determinant, because $SK_1(S)$ is already a discrete group. Hence
$\pi_0K_1(C_{\bf C}(B))\cong K^{-1}(B)$ and $\pi_0K_1(C_R(B))\cong KR^{-1}(B)$. Using the methods
outlined above, this goes through for $R_3$ also, giving $\pi_0K_1(C_Q(B))\cong KQ^{-1}(B)$.
In summary, the polynomially-generated vector bundles with an automorphism are classified up to 
homotopy by $\pi_0 K_1(R_i)$ above, and do not yield any nonzero element of $SK_1(C_i)$, just as 
polynomially-generated vector bundles do not yield
any nonzero element of $\widetilde{K}_0(C_i)$, the non-trivial part of $K_0(C_i)$, which was the statement 
of the no-go theorem for vector bundles.

The remarks that follow in this paragraph will not be used in the remainder of the paper.
As regards the second step, Milnor also mentions that, for the ring $C_{\bf C}(B)$, there is an exact
sequence involving the first few algebraic groups $K_j(C_{\bf C}(B))$ and the topological groups
$K^{-j}(B)$.
This exact sequence can be understood conceptually, and extended to all values of $j$, because in some of
Quillen's definitions of algebraic $K$-theory [e.g.\ the $B(S^{-1}S)$ construction \cite{weibelK}]
there is the option, in the case of rings such as $\bf C$, $R_1$, or $C_{\bf C}(B)$, of using two
different topologies under either of which the ring operations are continuous maps. The usual algebraic
theory corresponds in Quillen's treatment to using the discrete topology on the ring, while there is also
a natural ``continuous'' (non-discrete) topology (as in Milnor's discussion), the use of which in the
appropriate construction leads to the topological $K$ groups (\cite{weibelK}, sections IV.3.9, IV.4.12.3).
There is a ``change of topology'' functor (see the same references) that, for the given ring, leads to
an infinite long exact sequence involving the two types of $K$ theory and some relative groups, and which
explains Milnor's statements. (The references state that there is a map of spaces associated to the two
topologies on the ring; the $K$-groups are homotopy groups of these spaces, and the long exact sequence
for homotopy groups of two spaces with a map between them produces the long exact sequence of $K$ groups.)
We used a purely algebraic formulation, but then referred to a natural continuous topology when
identifying the connected component of the identity, before taking the homotopy sets; together, these
correspond to the change-of-topology functor on $R_1$.
Thus essentially the two steps we used are the composite of the change of rings and change of topology
to pass from $R_1$ (with the discrete topology) to $C_{\bf C}(T^d)$ (with the continuous
topology) and relate algebraic to topological $K$-theory; the functors can be considered in either order.
(Naturally, similar statements apply for the other rings
and other $K$-groups used in this paper.) We will continue to use the simpler homotopy-set point of view.


\section{Altland-Zirnbauer classes D, DIII, C, CI: relative $K_0$}
\label{sec:az}

In this section we deal with the final four symmetry classes, those characteristic of paired states of
fermions, the Bogoliubov-de Gennes or Altland-Zirnbauer (AZ) \cite{az} symmetry classes D, DIII, C, and CI.
(Other symmetry classes, such as BDI and CII, can also arise in relation to paired states, but have
already been covered.)

\subsection{AZ symmetry classes}

In this subsection, similar to the previous two groups of symmetry classes, we characterize each class,
and define the problem that needs to be solved algebraically, before turning to the methods to do so in
the following subsection. We will be able to treat the four cases in parallel, but we present details
in only one case, as an example.

We begin with what we view as the most basic class, class D. It is necessary to begin with a second
quantized ``reduced'' Hamiltonian $H_{\rm red}$, which has the most general form
\be%
H_{\rm red}=\frac{1}{2}C^\dagger
\left(\begin{array}{cc}h&\Delta\\
                   -\overline{\Delta}&-\overline{h}
                   \end{array}\right)C
 =\frac{1}{2}C^\dagger{\cal H}C,
    \ee%
where $C$ stands for a column vector of creation and destruction operators, in which the
first $M$ components are $c_\alpha$, the remaining $M$ are
$c_\alpha^\dagger$; further, for the $M\times M$ matrices $h$ and $\Delta$,
$h$ must be Hermitian  and $\Delta$ must be antisymmetric. Using the $2M\times 2M$ matrix $\cal H$ viewed
as acting in a tensor product of $M$ dimensional vector space and a two-dimensional space (``Nambu
space''), the required behavior of $\cal H$ can be characterized by
\be
\Sigma_x \overline{\cal H}\Sigma_x = - {\cal H},
\ee
as well as ${\cal H}^\dagger={\cal H}$. The operation on the left can also be described using
a time-reversal-like antilinear operator, $\tau=K\Sigma_x$ with $\tau^2=+I$, as $\tau{\cal H}\tau^{-1}$.
(Note we are {\em not}
saying the system has time-reversal symmetry here). Such $\cal H$ can be viewed as elements of
the Lie algebra of $O(2M)$, and have eigenvalues in $\pm E$ pairs; an orthonormal set of eigenvectors can
be assembled into an orthogonal matrix. The eigenvectors corresponding to a pair of eigenvalues $\pm E$ are
related by $\tau$, so have the form $w$, $\tau w$. The more familiar basis for $O(2M)$ can be obtained by
taking
the real and imaginary components of vectors; for the operators $c_\alpha$, $c_\alpha^\dagger$, they are
written as combinations of the now-familiar self-adjoint or ``Majorana'' operators. In this basis,
$\cal H$ becomes $i$ times a real antisymmetric matrix. The eigenvectors can be viewed as defining a
complex structure on the real vector space (i.e.\ choosing a real $2M\times 2M$ matrix $J$ with $J^2=-I$),
turning ${\bf R}^{2M}$ into ${\bf C}^M$; the choice of such a
complex structure can be labeled uniquely by a point in the space $O(2M)/U(M)$, because the
complex structure is invariant under a group isomorphic to $U(M)$ (a change of basis on ${\bf C}^M$).
In this point of view, the familiar Bogoliubov transformation (used to diagonalize a second-quantized
Hamiltonian in the form above) corresponds to a {\em change} from the given reference complex structure
to the equivalence class that contains the basis in which the single-particle Hamiltonian $\cal H$ is
diagonal.

For a translation-invariant system with $n$ orbitals per site, we obtain a $2n\times 2n$ Hamiltonian
matrix ${\cal H}_\bk$ in $\bk$ space, which obeys
\be
\tau{\cal H}_\bk \tau^{-1}=-{\cal H}_{-\bk}
\ee
with $\tau=K\Sigma_x$ as before, and also ${\cal H}_\bk^\dagger={\cal H}_\bk$. The explicit form can
be written as
\be%
{\cal H}_\bk=
\left(\begin{array}{cc}h_\bk&\Delta_\bk\\
                   -\overline{\Delta}_{-\bk}&-\overline{h}_{-\bk}
                   \end{array}\right).
    \ee%
Then the eigenvectors of ${\cal H}_\bk$
come in pairs $w_\bk=(u_\bk,v_\bk)^T$, $\tau w_{-\bk}=(\overline{v}_{-\bk},\overline{u}_{-\bk})^T$,
with eigenvalues $E_\bk$, $-E_{-\bk}$. Using an orthonormal set, we assemble these into a matrix
\be
W_\bk=\left(\begin{array}{cc}u_\bk&\overline{v}_{-\bk}\\v_\bk&\overline{u}_{-\bk}\end{array}\right)
\ee
(where here $u_\bk$, $v_\bk$ stand for $n\times n$ matrices) which obeys $\tau W_\bk \tau^{-1}=W_{-\bk}$
and $W_\bk^\dagger =W_\bk^{-1}$; $W_\bk$ represents the Bogoliubov
transformation at each $\bk$. We might describe this as representing a choice of a complex structure
on a Real vector bundle; it is similar to the Quaternionic case, class AII, except that
the energies come in $E$, $-E$ pairs, more like the chiral classes. At $\bk$ such
that $-\bk\equiv \bk$, it reduces to a complex structure on the real vector space, and
$W_\bk$ becomes an element of $O(2n)$.

For Wannier-type functions, we have only to find a set of such pairs $w_\bk$, $\tau w_{-\bk}$, which one
may think of as a set of wavefunctions for quasiparticle creation and annihilation operators, with of
course the overcompleteness property in $\bk$ space to make them Wannier-type. For compactly-supported
functions, they should in addition have components that are Laurent polynomials in $X_\mu=e^{ik_\mu}$.
For the latter, we have simply vectors $w$ (with entries in $R_1$) and its partner $\tau w$, in which
complex coefficients are conjugated and $X$ is unaltered [corresponding to
$X_\mu(\bk)=\overline{X_\mu(-\bk)}$].
There is a matrix $W_\bk$ with $\tau W_\bk \tau^{-1}=W_{-\bk}$ as above, and the vectors $w$ ($\tau w$)
together span the same space spanned by the first (second) $n$ columns of $W_\bk$ when evaluated at
$\bk$. (This does not necessarily mean that $W$ has polynomial entries.) $W$ represents the
transformation from the standard complex structure (corresponding to $W=I_{2n}$) to another one.
The vectors $w$ can be linearly combined with one another using coefficients in $R_1$  (complex
polynomials), so they generate a module over $R_1$, while $w\pm \tau w$ give real and imaginary
parts, which can be linearly combined only using coefficients in $R_2$ (Real polynomials), so they form a
module over $R_2$; we recall that $R_2$ is a subring of $R_1$: $R_2\subseteq R_1$. Hence the complex
structure turns an $R_2$ module into an $R_1$ module.

For the remaining symmetry classes, the details are similar but more intricate, and in principle can be
found in the literature \cite{az,kitaev,srfl} (for vector bundles, not for modules over polynomials); note,
however, that AZ
mainly focused on the Hamiltonian, not on the vector space or bundle formed from the eigenfunctions.
The structure parallels that in class D, where it involved the natural embedding (or inclusion) of rings
${\bf R}\subseteq {\bf C}$ for $d=0$, or $R_2 \subseteq R_1$ for Laurent polynomials, and for modules
over these rings an additional structure turning an $R_1$ module into an $R_2$ module (i.e.\ the reverse
direction). We briefly recall the symmetries involved in the remaining classes, in addition to the $\tau$
symmetry: For class DIII, time-reversal symmetry $\widehat{T}^2=-I$ is present. For class C, the system
admits an SU(2) ``spin rotation'' symmetry, but not time-reversal symmetry. For class CI, time-reversal
symmetry is present as well as spin-rotation symmetry. We present the inclusions of rings, showing the
$d=0$ case as well as the Laurent polynomial rings:
\be
\begin{array}{lll}
\hbox{D:}&{\bf R}\subseteq {\bf C},&R_2 \subseteq R_1;\\
\hbox{DIII:}&{\bf C}\subseteq {\bf H},& R_1\subseteq R_3;\\
\hbox{C:}&{\bf H}\subseteq M_2({\bf C}),&R_3\subseteq M_2(R_1);\\
\hbox{CI:}&{\bf C}\subseteq M_2({\bf R}),& R_1\subseteq M_2(R_2).
\end{array}
\label{RinS}
\ee
(Here and below $M_n(R)$ for a ring $R$ means the ring of $n\times n$ matrices with entries in $R$.)
The forms of all these inclusions can be understood using constructions of the complex numbers and
quaternions as matrices, as we have discussed; for example, $\bf C$ can be represented by $2\times 2$
real matrices in $M_2({\bf R})$, or as a subset of the quaternions $\bf H$, by representing $i$ as
${\hatj}=i\sigma_y$ (which is real) in the construction discussed earlier. (Similar forms apply
also for the rings $C_1$, $C_2$, $C_3$ of continuous functions on $T^d$, corresponding to
the rings $R_1$, $R_2$, $R_3$, respectively.) For use below, we also point out that in each of these
inclusions $R\subseteq S$, $S$ is a free module of rank $2$ over $R$ (generated over $R$ by $1$ and
$i$, $\hati$, $i1$, or $i{\hati}$, respectively, where the last three refer to the $2\times 2$ matrix
constructions). The remaining spaces that describe the spaces of possible structures in the general
(or $d=0$) case, corresponding to $O(2M)/U(M)$ for class D, are $U(2M)/Sp(2M)$, $Sp(2M)/U(M)$, and
$U(M)/O(M)$, respectively, in correspondence with the inclusions above. (The full list of ten spaces,
all related to topological $K$-theory, appears e.g.\ in Refs.\ \cite{karoubi,milnor}, as well as in Ref.\
\cite{az} in a slightly different way.)

\subsection{Classification by relative $K_0$}

For the analysis of the band structures in the AZ classes, we need to characterize the possible ways in
which a module over a ring $R$ can be extended to obtain a module over a ring $S$, where $R\to S$ is
an inclusion of rings, so $R\subseteq S$. To describe this, we first notice that for any $S$-module $M_S$,
there is an $R$-module $M_R$, obtained using the pullback or forgetful functor (see Sec.\ 
\ref{background}): as $M_S$ is an $S$-module,
it is certainly an $R$-module, when $R$ is viewed as a subset of $S$. (The subscript $R$ or $S$ records
the ring for which $M$ is viewed as a right module.) Different $S$-module structures can
be obtained from a reference one by following the map by an automorphism of $M_R$
as an $R$-module, while automorphisms of $M_S$ correspond to the same $S$-module structure; note that such
an automorphism maps to an automorphism of $M_R$. As automorphisms of projective modules $M_R$,
$M_S$ are described by $K_1(R)$ and $K_1(S)$, we expect that the desired classification should involve
the quotient of $K_1(R)$ by (the image under a homomorphism of) $K_1(S)$, though this quotient might not
exhaust the classification. For the example of class $D$,
in the simplest example of matrices or in zero-dimensional space, $R\subseteq S$ is ${\bf R}\subseteq
{\bf C}$, and the description just given appears to be an algebraic analog of the classifying space
$O/U$ mentioned above in similar terms, because $O(2M)$ [$U(M)$] describes the automorphisms of the
free $R$-modules of rank $2M$ (free $S$-modules of rank $M$)---that is, of real (complex) vector spaces.
We return to the precise $K$-theoretic characterization of the classification that we need after
introducing the correct machinery.

For the formal description in $K$-theory, we need the relative $K_0$ group associated to a functor
$\varphi$ that takes a category of $S$ modules to a category of $R$ modules; for various versions of
this, see Refs.\ \cite{rosenberg}
(page 131), \cite{karoubi} (section II.2.13), or \cite{weibelK} (section II.2.10), which is
the simplest. First, a functor is said to be {\em exact} if it maps a short exact sequence in the
first category to a short exact sequence in the second (thus, the categories must possess exact
sequences). The pullback functor of an inclusion is exact on the categories of
f.g\ modules of the two rings whenever $S$ is f.g.\ as an $R$-module, which is true for the examples
here. It also induces an exact functor between the categories of f.g.\ projective modules, provided $S$
is f.g.\ projective as an $R$-module; in our examples, $S$ is actually free of rank $2$. These general
statements follow
by representing the pullback functor as the tensor product $-\otimes S$, described in Sec.\
\ref{background}; see Ref.\ \cite{weibelK}, page 350. Then the definition of $K_0(\varphi)$ goes as
follows: we take
triples $(M_1,M_2,\alpha)$, where $M_1$, $M_2$ are $S$-modules in the category in question, and $\alpha$
is an isomorphism $\alpha:\varphi(M_1)\to\varphi(M_2)$ from the image of $M_1$ to the image of
$M_2$ under the functor $\varphi$; in other words, $M_1$ and $M_2$ become isomorphic after forgetting
the $S$-module structure. Then with some natural-looking equivalence relations imposed
on these triples (for which we defer to the references), we obtain a Grothendieck group, and when
the categories of modules are those of f.g.\ projective modules, we will denote it by $K_0(\varphi)$;
again we will write $[(M_1,M_2,\alpha)]$ for the equivalence class of a triple $(M_1,M_2,\alpha)$
in $K_0(\varphi)$. There is a natural map $K_0(\varphi)\to K_0(S)$, given by $[(M_1,M_2,\alpha)]\to
[M_1]-[M_2]$. For the inclusions $R\to S$ in our examples, there are natural homomorphisms
(called ``transfer maps'') $K_j(S)\to K_j(R)$ for $j=0$, $1$, which come directly from the forgetful
(pullback) functor. Clearly the image of $[(M_1,M_2,\alpha)]$ in $K_0(R)$ under the composite of these
maps is zero, because of the isomorphism $\alpha$.

When the functor $\varphi$ is also ``cofinal'' (or ``quasi-surjective''), essentially meaning that it
maps f.g.\ free modules to f.g.\ free modules, as is the case for the pullback in our examples (again
because $S$ is free of rank $2$), then there is an exact sequence of $K$-groups (Ref.\ \cite{karoubi}
sections II.2.20, II.3.22, or Ref.\ \cite{weibelK}, page 210),
\be
K_1(S)\to K_1(R)\stackrel{\partial}\to K_0(\varphi)\to K_0(S)\to K_0(R)
\label{ex_relative}
\ee
($\partial$ is the ``connecting map'').
If the image of $K_0(\varphi)$ in $K_0(S)$ is zero, then this shows that $K_0(\varphi)\cong K_1(R)/{\rm
im}\,K_1(S)$, as we anticipated above. If one knows the $K$-groups of the two rings, and the maps
between them in the exact sequence, then the sequence can be used to calculate $K_0(\varphi)$. The same
can also be done for the rings of continuous functions. We use a change of rings to an inclusion of rings of
(possibly matrices over) continuous functions that correspond to the inclusion $R\subseteq S$, and denote
it as $R'\subseteq S'$ [coming from replacing each $R_i$ by $C_i$ in the inclusions (\ref{RinS})], and
the corresponding pullback functor as $\varphi'$, leading to the calculation of $K_0(\varphi')$.
Finally, this can be related to the homotopy classification
of vector bundles at the end, as for the $K_0$, $K_1$ cases.

To cement the identification of the relative $K_0(\varphi)$ [or $K_0(\varphi')$, likewise] group as
the correct classification for the f.g.\ modules or bundles in the AZ symmetry classes (at least when
the modules are projective), we first note
that when $\varphi$ is cofinal, any class in $K_0(\varphi)$ can in fact be represented by the class of a
triple $[(M_1,S^n,\alpha)]$ for some $n$, so while $M_1$ is projective, $M_2=S^n$ is now free (see Ref.\
\cite{weibelK}, page 80; the proof is straightforward).
Since $\varphi(S^n)$ is free, this means that $\varphi(M_1)$ is isomorphic to a free $R$-module, which is
exactly the situation in the paired states in tight-binding models that we study, at least for the
projective modules over the polynomial rings (in view of the projective modules being stable free,
so possibly after taking direct sum with a free module), and for bundles: namely, when the pairing is
ignored, the system just becomes the tight-binding model band structure, which is trivial as a vector
bundle (i.e.\ free as a module over the ring of continuous functions corresponding to $R$). We note that
this description contains, but
is more general than, the description above as a quotient $K_1(R)/{\rm im}\,K_1(S)$,
since it allows ${\rm im}\,K_0(\varphi)\subseteq K_0(S)$ to be non-zero; instances of this occur for
$K_0(\varphi')$  for modules over the continuous functions (or for band structures) in class D in two
dimensions (and a calculation then leads to the correct results for the classification). We will see that
the image ${\rm im}\,K_0(\varphi)$ is always zero in the cases of the polynomial rings considered below.

We will now carry out the calculation in the four cases of interest. First, for class D, the exact
sequence reads
\be
K_1(R_1)\to K_1(R_2)\to K_0(\varphi)\to K_0(R_1)\to K_0(R_2).
\ee
For the polynomial rings, the $K_0$ groups are always $\cong \bf Z$, while the $K_1$ groups have
been determined earlier. Now we require information about the maps (homomorphisms of Abelian groups)
in the sequence. In general, it is sufficient to understand how the functor (pullback, in our case) acts
on some representative module in each equivalence class in each $K_0(S)$ and $K_1(S)$ group. In the
present case, the rings are Laurent polynomial extensions of division rings, and we know from the
earlier analysis that free modules of rank $m$ over each ring give representatives for the classes.
Thus in the present case, the pullback functor $\varphi$ maps the space of complex vectors (with polynomial
entries in $R_1$) of rank $m$ over $R_1$ to a space of Real vectors, by taking Real and ``Imaginary''
parts, thus producing a module of rank $2m$ over $R_2$. In this way, only free modules over $R_2$ of
even rank can be produced, so the last map in the sequence is multiplication by $2$, or $\times 2$,
on the integers,
and hence its kernel is zero. Thus $K_0(\varphi)$ maps to zero, and the map before it must be a
surjection. For $K_1$s, the relevant information is contained in the determinants (again using free
modules). Invertible matrices of size $m$ over $R_1$ map to invertible matrices of size $2m$
over $R_2$ when complex numbers are represented as $2\times 2$ real matrices, and the (real) determinant
of the latter is the square of the absolute value of the (complex) determinant of the former (in the
absolute value, the $X_\mu$s are treated simply as indeterminates, so not complex conjugated). Hence
the units $c\prod_\mu X_\mu^{m_\mu}$ in $R_1$ (which
are the possible values of the complex determinant) map to units $|c|^2\prod_\mu X_\mu^{2m_\mu}$, and
the first map is $|.|^2\oplus d.(\times 2)$ (i.e.\ absolute-value squared on ${\bf C}^\times$, and
multiplication by $2$ on each group of integers), mapping $K_1(R_1)={\bf
C}^\times\oplus d.{\bf Z}$ into $K_1(R_2)={\bf R}^\times\oplus d.{\bf Z}$. We can summarize these
statements by writing out the exact sequence explicitly as
\be
{\bf C}^\times\oplus d.{\bf Z} \stackrel{|.|^2\oplus d.{\times 2}}\to {\bf R}^\times\oplus d.{\bf Z}
\to K_0(\varphi)
\stackrel{0}\to {\bf Z} \stackrel{\times 2}\to {\bf Z}
\ee
(the unidentified connecting map being the quotient, which is a surjection).
Because ${\bf R}^\times$ contains negative as well as positive
real numbers, we find that the quotient of $K_1(R_2)$ by the image of the first map is therefore
\be
K_0(\varphi)=(d+1).{\bf Z}/2.
\ee
The results for $d=0$, $1$ agree with the topological $K$-group for this class, which is
$KR^{-2}(T^d)$ \cite{kitaev}; in this case there was no continuous part to divide out on passing from
the algebraic group for Real functions to the topological $K$-theory group.

The calculation for class DIII is similar. The exact sequence is
\be
K_1(R_3)\to K_1(R_1)\to K_0(\varphi)\to K_0(R_3)\to K_0(R_1).
\ee
Again, the pullback from matrices with Quaternion polynomial entries to matrices with complex entries
doubles the rank of a free module (over the respective rings), and maps of determinants behave similarly
as before. The sequence becomes
\be
{\bf R}^\times_{>0}\oplus d.{\bf Z} \stackrel{|.|^2\oplus d.{\times 2}}\to {\bf C}^\times\oplus d.{\bf Z}
\to K_0(\varphi)
\stackrel{0}\to {\bf Z} \stackrel{\times 2}\to {\bf Z}.
\ee
Hence the relative $K_0$ group for class
DIII is
\be
K_0(\varphi)= U(1)\oplus d.{\bf Z}/2,
\ee
and contains a continuous summand $U(1)$ (the analog of the additional ${\bf Z}/2$ in the class D case).
Since $U(1)$ is connected, it disappears in $\pi_0K_0(\varphi)$, the quotient by the connected component 
of the identity, which we will compare with the topological
$K$ group $KR^{-3}(T^d)$ (in particular, these agree for $d=0$ and $d=1$).

The case of class C involves the pullback from a matrix ring $S=M_2(R_1)$ to $R_3$. Algebraic $K$-theory
always exhibits invariance under Morita equivalence of rings, which means in particular that the
$K$-groups of a matrix ring are the same as those of the ring, that is $K_j(M_n(R))\cong K_j(R)$ for any
$j$ and any ring $R$. We therefore have
a functor from modules over $R_1$ to modules over $R_3$, and we note the inclusion $R_1\subseteq R_3$.
In $K$-theory, the original pullback functor is in fact Morita equivalent to the change-of-rings functor
corresponding to this inclusion. That is, for class C, we simply ``extend rings'' from $R_1$ to $R_3$ in
a natural way; this functor leaves the rank of a free module unchanged. The exact sequence of the pullback
is (Morita) equivalent to
\be
K_1(R_1)\to K_1(R_3)\to K_0(\varphi)\to K_0(R_1)\to K_0(R_3).
\ee
Working through the maps gives the sequence
\be
{\bf C}^\times\oplus d.{\bf Z} \stackrel{|.|\oplus d.{\rm id}}\to {\bf R}^\times_{>0}\oplus d.{\bf Z}
\stackrel{0}\to K_0(\varphi)
\stackrel{0}\to {\bf Z} \stackrel{\rm id}\to {\bf Z}
\ee
($\rm id:{\bf Z}\to{\bf Z}$ is the identity map on the integers), which gives, for class C,
\be
K_0(\varphi)=0.
\ee
For $d=0$, $1$, this agrees with the topological $K$-group for class C, $KR^{-6}(T^d)$.

Similarly for class CI, by Morita invariance the exact sequence is
\be
K_1(R_2)\to K_1(R_1)\to K_0(\varphi)\to K_0(R_2)\to K_0(R_1),
\ee
and the functor becomes the change-of-rings functor for $R_2\subseteq R_1$, which again leaves the rank of
a free module invariant. The details of the maps give
\be
{\bf R}^\times\oplus d.{\bf Z} \stackrel{i_*\oplus d.{\rm id}}\to {\bf C}^\times\oplus d.{\bf Z}
\to K_0(\varphi)
\stackrel{0}\to {\bf Z} \stackrel{\rm id}\to {\bf Z},
\ee
where $i_*:{\bf R}^\times \to{\bf C}^\times$ is the map induced from the inclusion $i:{\bf R}\to {\bf C}$
of the rings of real into complex numbers. This then implies that, for class CI,
\be
K_0(\varphi)= U(1).
\ee
[This $U(1)$ is the multiplicative group of complex numbers of absolute value $1$, modulo the subgroup
${\bf Z}/2=\{\pm 1\}$.] The homotopy group is $\pi_0K_0(\varphi) = 0$, and for $d=0$ and $1$ this is the
same as the topological $K$-group $KR^{-7}(T^d)$.

So far, we discussed and calculated the relative $K_0$ groups, which are defined using the categories of
f.g.\ projective modules over our rings. As in the earlier sections for other classes, what we actually
need for full generality is an analysis for more general modules, and it may be helpful to have a
corresponding result for relative $G_0$ groups, defined using the categories of all f.g.\ modules,
as an ``upper bound'' on the classification of the modules of interest,
even though in fact we do not strictly need the bound, as we will see in the following paragraph. A full
discussion for these relative $G_0$ groups is harder to find than for $K_0$ groups, however, the
condition for the
underlying pullback functor to be exact has already been given, and leads to maps $G_j(S)\to G_j(R)$
(see Ref.\ \cite{weibelK}, page 350). A high-level argument for the definition of
the relative group $G_0(\varphi)$ and the exact sequence can be obtained in higher algebraic $K$-theory
from Quillen's $Q$ construction, in which the absolute $K_j(R)$ [$G_j(R)$] groups are defined as
homotopy groups of certain spaces constructed from the categories of f.g.\ projective modules (f.g.\
modules), and for any exact functor between such categories (say, for two rings $R$ and $S$) there is a
corresponding map of the spaces (see again Ref.\ \cite{weibelK}, page 350). Then an exact sequence
containing relative groups $K_0(\varphi)$ [$G_0(\varphi)$] groups follows from the exact homotopy sequence
associated to the map. Finally, because we know that $G_j(R)\cong K_j(R)$, $G_j(S)\cong K_j(S)$, and also
the transfer maps induced from the pullback agree (\cite{weibelK}, page 425),
it follows that $G_0(\varphi)\cong K_0(\varphi)$ for the pullback functors $\varphi$, by using the exact
sequence again, and so the results for $G_0(\varphi)$ are the same as those already calculated above as
$K_0(\varphi)$.

Finally, we need to consider the relation of the modules over polynomial rings with the topological
classification in terms of vector bundles. This is much like the analysis in earlier sections, especially
the case of chiral symmetry classes that involved $K_1$. We carry it out here in a form that avoids any use
of $G_0(\varphi)$ for the polynomial rings. We will use the change of rings to the inclusions $R'\subseteq
S'$, and the corresponding pullback functors $\varphi'$, as already defined. First, an example in one of
these symmetry classes means a set of polynomial sections that generate a module $M$ over the ring $S$
[see the inclusions $R\subseteq S$ in (\ref{RinS})], with the
overcompleteness property that, after the change of rings to $S'$, we obtain an $S'$-module $M'$ with
generators that span the fibre of the
bundle (i.e.\ the space of states of the tight-binding model) at all $\bk$; by applying $\varphi'$,
we obtain a set of generators for an $R'$ module which is free (because it is the trivial bundle in
the tight-binding model viewed as an $R'$-module) and of the form $\varphi'(S'^n)$ for some $n$ (for the
same reason); of course the generators span the fibres of the corresponding bundle.

By the syzygy theorem,
$M$ has a projective resolution of finite length, and applying the pullback functor produces a projective
resolution of the pullback $R$-module $\varphi(M)$ also. Now we change rings to $R'\subseteq S'$ using the
change-of-rings functor, and apply it to the projective resolution as in earlier sections. Because we
assume that our polynomial sections have the overcompleteness property, the resulting sequence is an exact
sequence of projective $S'$-modules (indeed, free modules except possibly for $M'$ and at the $d$th place),
that is a projective resolution of the projective $S'$ module $M'$, and we also obtain another resolution
of the projective $R'$ module, the pullback $\varphi'(M')$. The pullback $\varphi'(M')$ is isomorphic to
a free module (trivial bundle) $\varphi'(S'^n)$ as already mentioned, via an isomorphism we call $\alpha'$.
In addition, we can assume that there are compatible maps $\alpha_i'$ making the pullbacks of the free $S'$
modules in the resolution isomorphic to free modules $\varphi'(S'^{n_i})$. These structures allow us to
apply $K_0(\varphi')$ to these exact sequences of triples, and we take it as given that
$K_0(\varphi')$ classifies the physically-relevant structure of the bundles in these symmetry classes
[i.e.\ equivalence classes of triples consisting of a pair of projective modules (i.e.\ bundles) and an
isomorphism]; see the discussion above. As we have an exact sequence (projective resolution), the
$K_0(\varphi')$ class for our triple $(M',\alpha',S'^n)$ containing our polynomially-generated bundle $M'$
is an alternating sum of those for the triples in the resolution. Those classes lie in the image of
$K_0(\varphi)$ in $K_0(\varphi')$ under the injective homomorphism induced from the change of rings. Hence
(passing finally to homotopy sets), the polynomially-generated bundles in these symmetry classes are
classified by elements of the $\pi_0 K_0(\varphi)$ groups of the polynomial rings which have already
been described. Again, these polynomially-generated bundles (which are non-trivial
only for classes D and DIII) can all be found in rank-one examples, by lifting the one-dimensional
constructions to higher dimensions by choosing some winding number in ${\bf Z}/2$ for each direction
(for class D in one dimension, the example is essentially the Kitaev chain \cite{KitaevChain}).
This concludes the constructions.

\section{Discussion}
\label{disc}

In this section, we discuss the general features of the results of this article. We have seen that, in
every one of the ten symmetry classes, the (stable) topological classification of vector bundles 
(or band structures) that are polynomially generated (i.e.\ can be constructed from compactly-supported 
Wannier-type functions; see section \ref{bunmod})
has a similar form: it can always be described as the classification (a group $0$, $\bf Z$, or ${\bf Z}/2$)
that arises for zero-dimensional systems (where the problem just reduces to matrices, essentially), plus
$d$ copies of the group (again $0$, $\bf Z$, or ${\bf Z}/2$) that classifies what can further arise in
that symmetry class in one space dimension within the general topological classification. (The result
for class A is the no-go theorem of DR \cite{dubail}.) Thus, all equivalence classes that can arise in
zero or one dimension can be obtained (of course, restricting functions in zero variables to be
polynomials has no effect, but should be included in the mathematical analysis). In higher dimensions,
the winding that can occur in one dimension can still occur in each of the $d$ directions, giving the $d$
copies mentioned; these are described as ``weak'' topological insulators or superconductors. But the other
invariants, including weak ones associated with dimensions $<d$ but $>1$, do not occur within polynomials.
This then constitutes the extension of the no-go theorem to other symmetry classes.

\begin{table}
\begin{tabular}{|c|c|c|c|c|c|c|}\hline
field&$p$& class & $\pi_0K_0(\varphi_p^{(d)})$& $d=0$& $d=1$ & $d=2$ \\ \hline
${\bf C}$&$0$&A&$\bf Z$&$\bf Z$&$\bf Z$&$2.{\bf Z}$\\ \cline{2-7}
&$1$&AIII&$d.{\bf Z}$&$0$&$\bf Z$&$2.{\bf Z}$\\ \hline
${\bf R}$&$0$&AI&$\bf Z$&$\bf Z$&$\bf Z$&$\bf Z$\\ \cline{2-7}
&$1$&BDI&${\bf Z}/2\oplus d.{\bf Z}$&${\bf Z}/2$&${\bf Z}/2\oplus {\bf Z}$&${\bf Z}/2\oplus
              2.{\bf Z}$\\\cline{2-7}
&$2$&D&$(d+1).{\bf Z}/2$&${\bf Z}/2$&$2.{\bf Z}/2$&$3.{\bf Z}/2\oplus{\bf Z}$\\ \cline{2-7}
&$3$&DIII&$d.{\bf Z}/2$&$0$&${\bf Z}/2$&$3.{\bf Z}/2$\\ \cline{2-7}
&$4$&AII&$\bf Z$&$\bf Z$&$\bf Z$&${\bf Z}/2\oplus{\bf Z}$\\ \cline{2-7}
&$5$&CII&$d.{\bf Z}$&$0$&${\bf Z}$&$2.{\bf Z}$\\ \cline{2-7}
&$6$&C&$0$&$0$&$0$&$\bf Z$\\ \cline{2-7}
&$7$&CI&$0$&$0$&$0$&$0$\\
 \hline
\end{tabular}
\caption{Table of results for topological phases that can be realized using compactly-supported
Wannier functions (polynomial sections)
or TNSs. First three columns: labels for symmetry classes of topological
phases. Fourth column: results of the analysis of the present paper for what can
be realized with polynomial sections in dimension $d$, up to homotopy. Fifth through seventh columns:
topological phases
in general non-interacting systems in dimensions $d=0$, $1$, and $2$, classified by
$K^{-p}(T^d)$ (for $\bf C$) or $KR^{-p}(T^d)$ (for $\bf R$), for comparison with the fourth
column.}\label{table_res}
\end{table}

The results are tabulated in Table \ref{table_res}. In this table we have labeled the classes by an
integer $p$ for both the real and complex classes, as well as with the Cartan symmetric space labels A,
AI, etc, used so far. The algebraic $K$ groups of the polynomial rings, modulo the connected component of
the identity, calculated earlier in this
paper have been denoted by $\pi_0K_0(\varphi_p^{(d)})$ in the table. For us, this is essentially only a
unified notation (explained further below). With this notation, the general result which was described
in the previous
paragraph, including both the complex and real cases, can be expressed as the following theorem
which encapsulates the results proved in this paper:\newline
{\it Theorem: polynomially-generated vector bundles are classified by the algebraic $K$-theory groups
\be
K_0(\varphi_p^{(d)})\cong K_0(\varphi_p^{(0)})\oplus d.[K_0(\varphi_p^{(1)})/K_0(\varphi_p^{(0)})],
\ee
for the symmetry classes labeled by $p=0$, $1$ (for $\bf C$), $p=0$, \ldots, $7$ (for $\bf R$), and for
all dimensions $d\geq 0$. The classification up to homotopy takes the same form, with $K_0$ replaced by
$\pi_0 K_0$ in each place; for these, the results coincide with the topological
$K$-theory groups $K^{-p}(T^d)$ (for $\bf C$) and $KR^{-p}(T^d)$ (for $\bf R$) for $d=0$, $1$ and
all $p$.}\newline
The mathematics behind this result is largely contained in the
so-called fundamental theorem of algebraic $K$-theory that we have mentioned, as well as the Hilbert
syzygy theorem (or the fact that the polynomial rings are regular) which was used repeatedly. We point
out that the zero-dimensional result $K_0(\varphi_p^{(0)})$ is always present as a summand here. 

In the Table, the columns labeled $d=0$ through $d=2$ contain the results of the topological
$K$-theory groups for the same symmetry classes in dimension $d$, that is $K^{-p}(T^d)$ for ${\bf C}$ and
$KR^{-p}(T^d)$ for $\bf R$. These columns are included for comparison with the results for
polynomial rings. Again, the zero-dimensional result, which is the same as
$K_0(\varphi_p^{(0)})$, is always present as a summand here. This summand can be viewed as classifying the
structure present at one point in the Brillouin torus (analogous to the other images of low-dimensional
groups, or weak invariants; the point should be one with $\bk\equiv -\bk$ in the real cases), or as a
``global'' invariant. In the literature, this part is frequently divided out or
omitted from the tabulated results, corresponding to the use of ``reduced'' $K$-groups, generically 
denoted $\widetilde{K}$. We believe that it is physically meaningful for the topological classification 
of band structures to retain it, that is to use unreduced $K$-groups, as it is the group of classes of 
a ``strong'' invariant
for $d=0$, and a ``weak'' invariant for $d>0$, which are on the same footing as the other 
invariants for $d>0$.

One sees in the Table that for five of the ten symmetry classes, namely, A, D, DIII, AII, and C,
there are topological phases that can occur in $d=2$ dimensions but cannot be realized with polynomials
or TNSs; these are the ``strong'' topological insulator or superconductor phases in two dimensions.
The difference becomes even larger in $d>2$ dimensions. In terms
of the results in Kitaev's paper \cite{kitaev}, our results are obtained by truncating the formula in
his eq.\ (26) to the terms $s=0$, $1$ only.

There is a unified way of describing results of topological $K$-theory for all symmetry classes
simultaneously, which can also be applied algebraically for modules over the rings of continuous
functions, as well as for rings of polynomials as used here. It extends the approach used for the AZ
classes here, viewing all of them as involving the extension of a Clifford algebra with $p$ generators
into another with $p+1$ generators, as shown in Karoubi's book, Ref.\ \cite{karoubi}, page 141.  Then
all the $K$ groups can be interpreted as relative $K_0$ groups for the appropriate pullback functor
$\varphi_p$ (hence the notation used in the Table). This viewpoint was used by Kitaev \cite{kitaev},
and has been popular in the physics literature. (While algebraic $K$-theory in general does not exhibit
Bott periodicity, that is, periodicity in $p$, the piece of it obtained by this method does, just like
the topological version.) We did not use this approach here because it requires
use of relative $K_0$ groups from the beginning. The more direct approach used here (which
requires no overt reference to a Hamiltonian) can also be employed for the $K$-theory of the rings
of continuous functions; that is essentially Karoubi's approach, except that, from the beginning, he
uses equivalence up to homotopy rather than up to isomorphism. We want to point
out that the use of Clifford algebras requires use of vector spaces whose dimension (or the dimension of
a tensor factor) is a sufficiently high power of $2$ (even higher if Dirac matrices are employed,
so that the Clifford algebra has a further $d$ generators). While this is not a problem when the goal is
to calculate $K$ groups of a given space $B$, or to construct examples in a given topological class, it
does not in general reflect the dimensions of the vector bundles that arise ``naturally'' or in band
theory. For our purposes, we wanted to consider the most general vector bundles we could, and so the
approach used herein seemed the most direct.

Finally we want to speculate on how an aspect of the results could generalize to interacting TNSs.
Certainly, the approach used here cannot be easily generalized to interacting systems. But the form of
the results may extend. There is an argument \cite{jerome} that
for lattice systems, topological phases that possess protected gapless edge excitations (like the
free-fermion topological phases considered in this paper) cannot be realized by a TNS with a gapped
parent Hamiltonian. The idea is that for a gapped TNS, correlations are short ranged, and then the
entanglement spectrum \cite{LiHaldane} (when the system is cut into two parts in position space)
is expected to be that of a short-range entanglement Hamiltonian acting in some suitable Hilbert space
of states confined close to the entanglement cut, and similar to that of a real edge. But for a TNS,
the rank of the entanglement Hamiltonian is bounded by the ``area'' of the cut times a constant related
to the rank of the tensors used
in the construction. It is then impossible for, say, a chiral spectrum to be obtained without the
Hamiltonian being non-analytic in $\bk$ space and therefore not short-ranged in position space. (Examples
for free-fermions can be seen e.g.\ in the figures in DR \cite{dubail}.) For a generic ground state
(not a TNS), this issue is avoided because the low-lying entanglement spectrum that resembles the edge
merges at higher pseudoenergies with a continuum coming from the bulk, which obviates the argument.

This argument does seem reasonable, but there is an additional point we wish to mention here: first,
the argument does not apply in one dimension; in that case, gapless edge modes are simply zero modes,
and there is no objection to them in a TNS on grounds of non-analyticity. Indeed, many examples are known
of one-dimensional topological phases that are TNSs (or MPSs) with gapped parent Hamiltonians: the Kitaev
chain is of this type. Then as for ``weak'' topological insulators and superconductors, there are phases
in higher dimensions that reflect (wholly or in part) topological phases from lower dimensions. (For
example, there can be higher-dimensional quantum Hall phases that are essentially two dimensional
quantum Hall states occurring in layers that are stacked, and without significant interaction between
them.) For behavior that results from a one-dimensional topological phase, the entanglement spectrum should
contain a large number of degenerate zero modes coming from the one-dimensional systems making it up. It is
possible for these to mix (that is, it is sometimes allowed by symmetries), which in some cases can split
the degeneracy of the modes (so they become non-zero-pseudoenergy modes). Such an entanglement spectrum,
for a phase that is derived from one-dimensional phases in each direction of space, is not forbidden by
the argument mentioned just now. This is reflected in the form of the generalized no-go theorem obtained
in this paper, and we expect that it is a general feature that occurs in interacting phases also. Likewise, 
we have viewed the zero-dimensional system as a single site, and there is no edge, and hence no gapless
edge modes. It makes sense that the corresponding weak invariants can appear in higher-dimensional 
polynomially-generated bundles, as we found.

\section{Conclusion}

The problem of compactly-supported Wannier functions, or polynomially-generated vector bundles, and
the results have been described both in the Introduction and in the preceding Discussion section, so we
will be brief here. The generalized form of the DR no-go theorem, proved in this paper, states that,
apart from the classification of zero-dimensional (or global) aspects of the band structures, for each of
the symmetry classes in the ``tenfold way'' classification for lattice models with only translational
symmetry, the only topological (stable) equivalence classes of vector bundles that can be obtained as
polynomially-generated bundles are those that have a ``winding number'' of a one-dimensional system in
the same symmetry class in each of the $d$ directions of space, and nothing else. The allowed possibilities
were listed explicitly in Table \ref{table_res}. These results imply
similar statements for free-fermion TNSs: any free-fermion TNS that gives rise to a bundle not in
the list will have only gapless parent Hamiltonians. They also apply to flat-band
Hamiltonians: For a strictly short-range flat-band Hamiltonian for a band structure (with the symmetries
assumed in this discussion) in any of the ten symmetry classes, and if the flat band is separated by a
gap from the remainder of the spectrum at all $\bk$, then the vector bundle of the flat band must be one
of those in the list.

The classification used here was in terms of algebraic $K$-theory, because of the rings of polynomial
functions that appeared in place of the more generic rings of continuous functions that appear in
connection with topological $K$-theory. It might be thought that the results can be described simply as
using the former in place of the latter, which results in the contrasting $K$ groups. There is a little
more to it, however, as in order to obtain full generality in the sets of Wannier-type functions, or
polynomial sections generating the bundle, it was necessary to venture beyond the projective modules,
which are classified directly by these groups. The syzygy theorem came to our rescue, allowing the bundles
in these more general cases nonetheless to be related to the classification by the $K$ groups.

It might now be interesting to extend the results to tight-binding models with additional crystallographic
symmetries, by using equivariant algebraic $K$-theory. The extension to
TNSs of interacting systems of fermions would also be interesting, but it is not clear
what techniques could be used to do this.

Finally, we want to emphasize that our results do not necessarily mean that other tensor-network
constructions, different from those considered here, for topologically non-trivial states cannot work.
As a concrete example, the scheme of Ref.\ \cite{BeriCooper} uses a tensor network to produce
approximate values for expectations of products of local operators in (non-TNS) trial topological states,
rather than using a TNS as a trial ground state as discussed here for free fermions. This alternative
approach does not appear to be affected by the results presented herein.

\acknowledgements

The author is grateful for the previous collaboration with J. Dubail and for discussions with
J. H\"{o}ller and A. Alexandradinata which stimulated this work; A. Alexandradinata is also thanked
for pointing out some of the references. J. H\"{o}ller, A. Alexandradinata, B. Bradlyn, C. Kane,
G. Moore, M. Cheng, and V. Ozoli\c{n}\v{s} are thanked for comments on a draft and for other remarks 
or discussions. The research was supported by NSF grant No.\ DMR-1408916.


\end{document}